\documentclass{emulateapj}
\shorttitle{Photoeccentric Effect}
\shortauthors{Dawson and Johnson}
\def\kep{\emph{Kepler\ }}
\def\prob{{\rm prob}}
\def\emin{e_{\rm min}}

\begin{document}
\title{The Photoeccentric Effect and proto hot Jupiters I. Measuring photometric eccentricities of individual transiting planets}
\slugcomment{Received 2012 April 5; accepted 2012 July 9; published 2012 August 21}
\author{Rebekah I. Dawson\altaffilmark{1}}
\affil{Harvard-Smithsonian Center for Astrophysics \\ 
60 Garden St, MS-10, Cambridge, MA 02138}
\author{John Asher Johnson}
\affil{Department of Astronomy, California Institute of Technology,\\ 
1200 East California Boulevard, MC 249-17, Pasadena, CA 91125, USA}

\affil{NASA Exoplanet Science Institute (NExScI), CIT Mail Code 100-22,\\ 
770 South Wilson Avenue, Pasadena, CA 91125}

\altaffiltext{1}{{\tt  rdawson@cfa.harvard.edu}}

\begin{abstract}
Exoplanet orbital eccentricities offer valuable clues about the history of planetary systems. Eccentric, Jupiter-sized planets are particularly interesting: they may link the ``cold'' Jupiters beyond the ice line to close-in hot Jupiters, which are unlikely to have formed in situ. To date, eccentricities of individual transiting planets primarily come from radial velocity measurements. \kep has discovered hundreds of transiting Jupiters spanning a range of periods, but the faintness of the host stars precludes radial velocity follow-up of most. Here we demonstrate a Bayesian method of measuring an individual planet's eccentricity solely from its transit light curve using prior knowledge of its host star's density. We show that eccentric Jupiters are readily identified by their short ingress/egress/total transit durations -- part of the ``photoeccentric'' light curve signature of a planet's eccentricity --- even with long-cadence \kep photometry and loosely-constrained stellar parameters. A Markov Chain Monte Carlo exploration of parameter posteriors naturally marginalizes over the periapse angle and automatically accounts for the transit probability. To demonstrate, we use three published transit light curves of HD~17156\,b to measure an eccentricity of $e = 0.71_{-0.09}^{+0.16}$, in good agreement with the discovery value $e = 0.67\pm0.08$ based on 33 radial-velocity measurements. We present two additional tests using actual \kep data. In each case the technique proves to be a viable method of measuring exoplanet eccentricities and their confidence intervals. Finally, we argue that this method is the most efficient, effective means of identifying the extremely eccentric, proto hot Jupiters predicted by Socrates et al. (2012).
\end{abstract}


\section{Introduction}

Many exoplanets have highly eccentric orbits, a trend that has been interpreted as a signature of the dynamical processes that shape the architectures of planetary systems \citep[e.g.][]{2008J,2008F,2011N}. \emph{Giant} planets on eccentric orbits are of particular interest because they may be relics of the same processes that created the enigmatic class of planets known as hot Jupiters: planets on very short period (P $<$ 10 days) orbits that, unlike smaller planets \citep[e.g.][]{2012H}, could not have formed in situ. Hot Jupiters may have smoothly migrated inward through the disk from which they formed \citep[e.g.][]{1980G,1997W,2005A,2008I,2011BK}. Alternatively, the typical hot Jupiter may have been perturbed by another body onto an eccentric orbit \citep[see][]{2012NF}, with a star-skirting periapse that became the parking spot for the planet as its orbit circularized through tidal dissipation, initiated by one of several perturbation mechanisms \citep[e.g.][]{2003W,2006F,2011W}. 

\citet{2012SK} (hereafter S12) refer to this process as ``high eccentricity migration" (HEM). If HEM were responsible for hot Jupiters, at any given time we would observe hot Jupiters that have undergone full tidal circularization, failed hot Jupiters that have tidal timescales too long to circularize over the star's lifetime, and proto hot Jupiters that are caught in the process of tidal circularization. S12 predicted that the \kep Mission should detect several ``super-eccentric'' proto hot Jupiters with eccentricities in excess of 0.9. This prediction was tested by \citet{2012D} on a sample of eclipsing binaries in the \kep field: in an incomplete search, they found 14 long-period, highly eccentric binaries and expect to eventually find a total of 100. 

As a test of planetary architecture theories, we are devoting a series papers to measuring the individual eccentricities of the \kep Jupiters to either identify or rule out the super-eccentric proto hot Jupiters predicted by S12. In this first paper, we describe and demonstrate our technique for measuring individual eccentricities from transit light curves. Measuring the eccentricity of a Jupiter-sized planet is also key to understanding its tidal history \citep[e.g.][]{2008Ja, 2010H} and tidal heating \citep[e.g][]{2007M,2008Jb}, climate variations \citep[e.g][]{2011K}, and the effect of the variation in insolation on the habitability \citep[e.g][]{2010SR,2010D} of possible orbiting rocky exomoons detectable by \kep \citep[e.g.][]{2009K}. 

To date, the measurements of eccentricities of individual transiting planets have been made through radial velocity follow-up, except when the planet exhibits transit timing variations \citep[e.g.][]{2012N}. However, a \emph{transit light curve} is significantly affected by a planet's eccentricity, particularly if the photometry is of high quality: we refer to the signature of a planet's eccentricity as  the ``photoeccentric'' effect. One aspect is the asymmetry between ingress and egress shapes \citep{2007BM,2008K}. The eccentricity also affects the timing, duration, and existence of secondary eclipses \citep{2009KV,2012D}. The most detectable aspect of the photoeccentric effect in \kep photometry for long-period, planet-sized companions is the transit event's duration at a given orbital period $P$, which is the focus of this work. 

Depending on the orientation of the planet's argument of periapse ($\omega$), the planet moves faster or slower during its transit than if it were on a circular orbit with the same orbital period (\citealt{2007B}, \citealt{2008B}, Ford, Quinn, and Veras 2008, hereafter FQV08; \citealt{2011M}). If the transit ingress and egress durations can be constrained, the duration aspect of the photoeccentric effect can be distinguished from the effect of the planet's impact parameter ($b$), because although $b > 0$ shortens the full transit duration ($T_{\rm 23}$, during which the full disk of the planet is inside the disk of the star, i.e. from second to third contact), it lengthens the ingress/egress duration. Therefore, with prior knowledge or assumptions of the stellar parameters, combined with measurements from the light curve of the planet's period and size ($R_P/R_\star$), one can identify highly eccentric planets as those moving at speeds inconsistent with a circular orbit as they pass in front of their stars (see also \S3 of \citealt{2007B},  \S3.1 of FQV08).

\citet{2007B} presented the first comprehensive description of the effects of orbital eccentricity on a transit light curve, including that a short transit duration corresponds to a minimum eccentricity, contingent on the measurement of $b$ and of the host star's density. \citet{2008B} discussed the effect of orbital eccentricity on transit detection and on the inferred distribution of planetary eccentricities. FQV08 laid out the framework for using photometry to measure both the distribution of exoplanet eccentricities and, for high signal-to-noise transits of stars with known parameters, the eccentricities of individual planets. They derived expressions linking the orbital eccentricity to the transit duration and presented predicted posterior distributions of eccentricity and $\omega$ for a given ratio of: 1) the measured total transit duration (i.e. from first to fourth contact, including ingress and egress) $T_{14}$ to 2) the $T_{14}$ expected for a planet on a circular orbit with the same $b$, stellar density $\rho_\star$, and $P$. Then they showed how the distribution of planetary transit durations reveals the underlying eccentricity distribution. FQV08 focused on the possibility of measuring the eccentricity distribution of terrestrial planets, which has implications for habitability. Here we will show that the technique they describe for measuring \emph{individual} planet eccentricities is particularly well-suited for Jupiter-sized planets.

The work of FQV08 was the basis for several recent analyses of high-precision light curves from the \kep mission that have revealed information about the eccentricity distribution of extra-solar planets and the eccentricities of planets in multi-transiting systems. By comparing the distribution of observed transit durations to the distribution derived from model populations of eccentric planets, \citet{2011M} ruled out extreme eccentricity distributions. They also identified individual planets with transit durations too long to be consistent with a circular orbit; these planets are either on eccentric orbits (transiting near apoapse) or orbit host stars whose stellar radii are significantly underestimated. 

\citet{2012KC} used the distribution of transit durations to determine that the eccentricity distribution of \kep planets matches that of planets detected by the RV method and to discover a trend that small planets have less eccentric orbits. In contrast, \citet{2012P} found that the distribution of eccentricities inferred from the transit durations is not in agreement with the eccentricity distribution of the RV sample; they suggested that the difference may be due to errors in the stellar parameters. Finally, \citet{2012K} presented a method that they refer to as \emph{Multibody Asterodensity Profiling} to constrain eccentricities of planets in systems in which multiple planets transit. They noted that one can also apply the technique to single transiting planets, but discouraged doing so, except for planets whose host star densities have been tightly constrained (e.g. by asteroseismology). FQV08 recommend measuring eccentricities photometrically only for planets with ``well-measured stellar properties'' but also point out the weak dependence of eccentricity on stellar density.

In this work we apply the idea first proposed by FQV08 to real data and demonstrate that we can measure the eccentricity of an individual transiting planet from its transit light curve. We show that this technique is particularly well-suited for our goal of identifying highly eccentric, giant planets. In \S2, we show that even a loose prior on the stellar density allows for a strong constraint on the planet's orbital eccentricity. In \S3, we argue that Markov Chain Monte Carlo (MCMC) exploration of the parameter posteriors naturally marginalizes over the periapse angle and automatically accounts for the transit probability. We include both a mathematical and practical framework for transforming the data and prior information into an eccentricity posterior. In \S4, we measure the eccentricity of HD~17156\,b from ground-based transit light curves alone, finding good agreement with the nominal value from RV measurements. We also measure the eccentricity of a transit signal injected into both short and long cadence \kep data and of \kep Object of Interest (KOI)~686.01 from long-cadence, publicly-available \kep data, finding an eccentricity of $e = 0.62_{-0.14}^{+0.18}$. In \S5, we present our program of ``distilling'' highly eccentric Jupiters from the KOI sample and we conclude (\S 6) with prospects for further applications of the photoeccentric effect.

\section{Precise eccentricities from loose constraints on stellar density}

To first order, a transiting planet's eccentricity and its host star's density depend degenerately on transit light 
	\begin{minipage}{7.0in}
\setlength{\parindent}{1cm}
\noindent 
curve observables. \citet{2012K} harnessed the power of multiple planets transiting the same host star to break this degeneracy \citep[see also][]{2010R}. Yet, as FQV08 first pointed out, although the transit observables depend on the stellar density, this dependence is weak (the ratio of the planet's semi-major axis to the stellar radius $a/R_\star \propto \rho_\star^{1/3}$). Thus a loose prior on the stellar density should allow for a strong constraint on the eccentricity.

In the limit of a constant star-planet distance during transit and a non-grazing transit (such that the transit is approximately centered at conjunction), \citet{2010K} derived the following expression (\citealt{2010K} Equations 30 and 31) for $T_{14}$, the duration from first to fourth contact (i.e. the total transit duration including ingress and egress), and for $T_{23}$, the duration from first to third contact (i.e. the full transit duration during which the full disk of the planet is inside the disk of the star):
\begin{equation}
\label{eqn:dur}
T_{14/23} = \frac{P}{\pi} \frac{(1-e^2)^{3/2}}{(1+e \sin \omega)^2} \arcsin\left[\frac{\sqrt{(1 +/- \delta^{1/2} )^2 -(a/R_\star)^2 (\frac{1-e^2}{1+ e \sin\omega})^2 \cos^2 i}}{ (a/R_\star) \frac{1-e^2}{1+ e \sin\omega} \sin i } \right]
\end{equation}
where $P$ is the orbital period; $e$ is the eccentricity; $\omega$ is the argument of periapse; $R_\star$ is the stellar radius; $\delta=(R_p/R_\star)^2$ is the fractional transit depth with $R_p$ the planetary radius; $a$ is the semi-major axis; and $i$ is the inclination.
\noindent By combining $T_{14}$ and $T_{23}$, we can rewrite Equation (\ref{eqn:dur}) as 
\begin{equation}
\label{eqn:durs}
\sin^2 (  \frac{\pi}{P} \frac{[1+ e \sin\omega ]^2}{(1-e^2)^{3/2}} T_{14}) - \sin^2 (  \frac{\pi}{P} \frac{[1+ e \sin\omega ]^2}{(1-e^2)^{3/2}} T_{23}) =   \frac{4  \delta^{1/2}  (1+ e \sin\omega)^2}{\sin^2 i \ (a/R_\star)^2 (1-e^2)^2 } 
\end{equation}
\noindent Using the small angle approximation, which is also used by \citet{2010K}, allows us to group the transit light curve observables on the right-hand side:
\begin{equation}
\label{eqn:areccsini}
\frac{a}{R_\star} g(e,\omega) \sin i= \frac{2 \delta^{1/4} P}{\pi \sqrt{T_{14}^2- T_{23}^2}}
\end{equation}
where
\begin{equation}
\label{eqn:g}
g(e,\omega)=\frac{1+e\sin\omega}{\sqrt{1-e^2}}
\end{equation}
The $g$ notation is inspired by \citet{2010K} and \citet{2012K}'s variable $\Psi$, for which $\Psi = g^3$. Dynamically, $g$ is the ratio of the planet's velocity during transit (approximated as being constant throughout the transit) to the speed expected of a planet with the same period but $e=0$. Note that $\omega$ is the angle of the periapse from the sky plane, such that $\omega=90^\circ$ corresponds to a transit at periapse and $\omega=-90^\circ$ to a transit at apoapse. For a given $P$ and $\delta$, $T_{14}$ and $T_{23}$ are shortest (longest) and $g$ largest (smallest) when the planet transits at periapse (apoapse). Moreover, if we approximate $\sin i = 1$, we can rewrite Equation (\ref{eqn:areccsini}) as:
\begin{equation}
\label{eqn:arecc}
\frac{a}{R_\star} g(e,\omega)= \frac{2 \delta^{1/4} P}{\pi \sqrt{T_{14}^2- T_{23}^2}}
\end{equation}
\noindent Finally, using Kepler's third law and assuming that the planet mass is much less than the stellar mass $(M_p \ll M_\star)$, the transit observables can be expressed in terms of the stellar density $\rho_\star$:
\begin{equation}
\label{eqn:rhoecc}
\rho_{\star}(e,\omega) = g(e,\omega)^{-3} \rho_{\rm circ}
\end{equation}
where
\begin{equation}
\label{eqn:rhocirc}
\rho_{\rm circ} =\rho_{\star}(e = 0 ) = \left[\frac{2 \delta^{1/4}}{\sqrt{T_{14}^2- T_{23}^2}}\right]^{3}\left(\frac{3P}{G\pi^2}\right)
\end{equation}
\noindent Although Equation~\ref{eqn:rhoecc} was derived under several stated approximations, the relationships among $\rho_\star$, $e$, and $\omega$ are key to understanding how and to what extent we can constrain a transiting planet's eccentricity using a full light curve model. Because $g(e,\omega)$ is raised to such a large power, a small range of $g(e,\omega)$ corresponds to a large range in the ratio $\rho_\star/\rho_{\rm circ}$, i.e. the ratio of the true stellar density to the density measured from fitting a circular transit light curve model.  For instance, the assumed value of $\rho_\star$ would need to be in error by two orders of magnitude to produce the same effect as a planet with $e = 0.9$ and $\omega = 90^\circ$. Thus the $\rho_{\rm circ}$ derived from the transit light curve strongly constrains $g$, even with a weak prior on $\rho_\star$, because $g \propto \rho_\star^{1/3}$.
\end{minipage}
\clearpage
\subsection{Constraints on $\rho_{\rm circ}$ from the light curve: common concerns}

One might worry that long-cadence data, such as the 30-minute binning of most \kep light curves, cannot resolve the ingress and egress times sufficiently to constrain $a/R_\star$, or equivalently $\rho_{\rm circ}$. In other words, one might worry that $a/R_\star$ is completely degenerate with $b$, and hence that the denominator of Equation (\ref{eqn:arecc}) is unconstrained. This is often the case for small planets. However, Jupiter-sized planets have high signal-to-noise transits and longer ingress and egress durations (due to the large size of the planet). See \S2.1 of FQV08 for an analysis of how the precision of \kep data affects constraints on the total, ingress, and egress durations.

Furthermore, even if the ingress is unresolved or poorly resolved, it is often impossible for the impact parameter $b$ to account for the short duration of a highly eccentric, Jupiter-sized planet's non-grazing transit. The maximum non-grazing impact parameter is $1-R_P/R_\star \lesssim 0.9$ for a Jupiter around a Sun-like star. Imagine that an eccentric planet transits at zero impact parameter (i.e. travels across $2R_p + 2R_\star$) at speed $g$. If we instead assume that planet is transiting at its circular speed $g = 1$ across the short chord of length ($2\sqrt{(R_\star+R_p)^2 - (b_{\rm large~enough} R_\star)^2}$), the required impact parameter would be:
\begin{equation}
b_{\rm large~enough} \approx (1+ \delta^{1/2})\sqrt{1-1/g^2}
\end{equation}
For $g = 2.38$ (corresponding to $e=0.7, \omega = 90^\circ$) and $\delta^{1/2} = R_p/R_\star = 0.1$, $b$ would need to be $\approx 0.998$, which would be inconsistent with a non-grazing transit. In contrast, a planet with $R_p/R_\star = 0.01$ would have $b_{\rm large~enough} \approx 0.917$, consistent with the $b < 0.99$ necessary for a non-grazing transit. We note this effect simply to highlight a constraint that arises naturally when fitting a \citet{2002M} transit model to a light curve.

Additionally, with a properly binned model \citep[as discussed in][who advocates resampling the data times, computing a model light curve, and then smoothing to match the data cadence]{2010Kb}, multiple transits allow for constraints on the ingress and egress, even if they are poorly resolved in a single transit. We demonstrate eccentricity measurements using long-cadence data in \S4.2.

Another concern regards the degeneracy of $a/R_\star$ and $b$ with the limb-darkening parameters. Limb darkening causes the shape of the transit to be rounded instead of flat, potentially causing confusion between the full transit and the ingress/egress. However, in practice we find that it makes little difference whether we freely vary the limb darkening parameters or impose a normal prior based on the stellar parameters (e.g. the coefficients computed for the \kep bandpass by \citealt{2010S}). FQV08 also find that limb darkening does not have a significant effect on the other parameters, as demonstrated through tests on simulated light curves (see FQV08 \S2.1 and FQV08 Figure 5).

Finally, one might worry about dilution by light from a nearby or background star blended with the target star (see \citealt{2011JA} for a \kep example). Dilution would cause $R_p/R_\star$ to appear too small. Consider the impact that dilution would have on the derived parameters of an eccentric planet transiting near periapse. The ingress and egress durations would be longer than expected, and the inferred maximum impact parameter to avoid a grazing orbit (i.e. $1-R_p/R_\star$) would be too large. Both of these effects would caused the planet's orbit to appear \emph{less} eccentric (or, equivalently, for $\rho_{\rm circ}$ to appear smaller; see \citealt{2010KT} for a formal derivation of the effect of blending on the measurement of $a/R_\star$). Therefore, dilution would not cause us to overestimate a planet's eccentricity, if the transit duration is shorter that circular. Moreover, because $\rho_{\rm circ}$ depends only weakly on the transit depth (Equation \ref{eqn:rhocirc}), the effect of blending on the eccentricity measurement is small. We quantify this effect through an example in the next subsection.

Furthermore, if we were to mistakenly attribute an apparently overly-long transit caused by blending to a planet transiting near apoapse, the resulting false eccentricity would be quite small. Imagine that the planet is on a circular orbit, but that the blend causes us to measure $\rho_{\rm circ}=(1-f) \rho_\star$, where $0 < f \ll 1 $. The inferred $g$ would be $g = [\rho_{\rm circ}/\rho_\star]^{1/3} \approx 1 - f/3$, very close to the true $g = 1$ of the circular orbit.

\subsection{Constraints on eccentricity}

From Equation~\ref{eqn:rhoecc}, it might appear that $e$ and $\omega$ are inextricably degenerate for a single transiting planet. Certainly, if $\rho_{\rm circ}$ is consistent with $\rho_\star$, any eccentricity is consistent with the transit observables. However, a nominal value of $\rho_\star$ smaller than $\rho_{\rm circ}$ translates to a \emph{minimum eccentricity} $e_{\rm min}$, the value obtained by assuming the planet transits at periapse ($\omega = 90^\circ$; see also \citealt{2007B}, \S3; \citealt{2012KC}, \S4). Conversely, a value of $\rho_\star$ larger than $\rho_{\rm circ}$ corresponds to an $e_{\rm min}$ obtained by assuming the planet transits at apoapse ($\omega = - 90^\circ$). Therefore, we can easily identify planets with large eccentricities. A full MCMC exploration provides a confidence interval that shrinks as $e \rightarrow 1$, as we discuss in detail in \S3. For example, consider a planet with an eccentricity of 0.9 that transits at semilatus rectum ($\omega=0$). Based on the transit light curve observables, we would deduce that it has an eccentricity of at least $e_{\rm min} = 0.68$. A planet transiting at semilatus rectum with $e = 0.98$ would have a deduced $e_{\rm min} = 0.92$. Above the sharp lower limit $e_{\rm min}$, the eccentricity posterior probability falls off gradually, as we discuss in \S3. Note that the $e_{\rm min}$ we have defined here, which assumes we can distinguish between $b$ and $\rho_{\rm circ}$ (i.e. via some constraint on ingress/egress time), is a stronger limit than the minimum eccentricity from the constraint that the transit be non-grazing (which we discussed in \S2.1).

Returning to the issue of contamination by blending (discussed in \S2.1), consider a transit with $g = 2.5$ and thus $e_{\rm min} = 0.724$. If the transit depth were diluted by a factor\footnote{This is a worst-case scenario because in fact we could easily detect a companion causing such a large dilution.} of 0.9 by an undetected second star in the photometric aperture, we would measure $g = 0.9^{1/4} 2.5 =2.435$ and infer nearly the same minimum eccentricity of $e_{\rm min} = 0.711$.  Finally, imagine that some of the constraint on $g$ measured from the light curve came from the non-grazing shape of the transit, implying an impact parameter greater than $1-R_p/R_\star$. If the $R_p/R_\star$ measured from the diluted transit curve were 0.1, the inferred maximum impact parameter would be 0.9. If the true $R_p/R_\star$ is $5\%$ larger, then the maximum impact parameter should be 0.895. This translates into a negligible effect on the constraint on $g$.

\begin{figure}[htbp]
\begin{centering}
\includegraphics{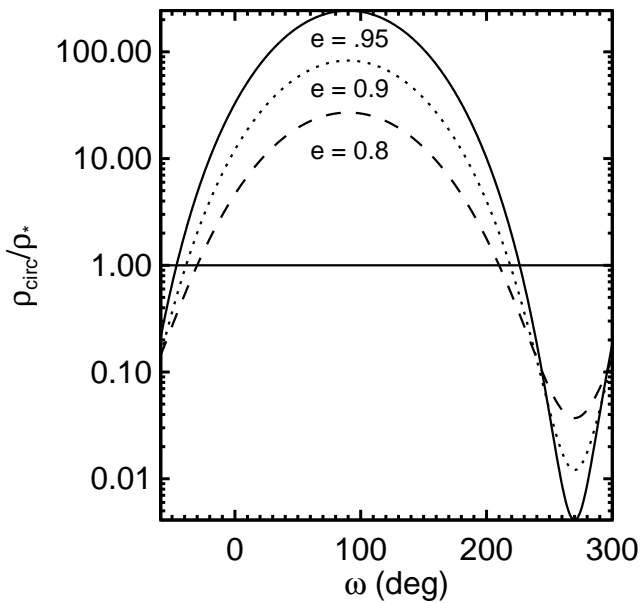}
\caption{The ratio of the circular density to the nominal stellar density, $\rho_{\rm circ}/\rho_\star$, required for a circular model to account for the transit observables of an eccentric planet. The ratio is plotted as a function of the planet's argument of periapse. The solid (dotted, dashed) line corresponds to a planet with an eccentricity of 0.95 (0.9, 0.8). For a large range of periapse angles, one would infer a density much larger than the nominal value if one modeled the eccentric planet's orbit as circular.\label{fig:rhom}}
\end{centering}
\end{figure}

In Figure \ref{fig:rhom}, we plot $\rho_{\rm circ}$ as a function of $\omega$. Centered at $\omega = 90^\circ$ is a broad range of $\omega$ for which $\rho_{\rm circ}$ would be quite high. For example, for $e = 0.9$, $\rho_{\rm circ}$ would be erroneously high by a factor of 10-100 for $-3^\circ < \omega < 183^\circ$, over half the possible orientations. Moreover, although the periapses of eccentric planets are intrinsically randomly oriented throughout the galaxy, based on geometry eccentric planets with $\omega \approx 90^\circ$ are more likely to transit.  For example, from a population of planets with $e = 0.9 \left( 0.95, 0.99 \right)$  and a given orbital separation, we would be able to observe 19 (39, 199) times as many transiting at periapse as at apoapse. 

Another happy coincidence is that the true stellar density is unlikely to be \emph{higher} than the \kep Input Catalog \citep[KIC,][]{2010B} value by a factor of 10. The opposite situation is common; a star identified as being on the main sequence may actually be a low-density subgiant or giant \citep[e.g.][Dressing et al. 2012, in prep]{2012MG}. Conversely, there are not many stars with the density of lead. Even when precise measurements of the stellar density are unavailable, our basic knowledge of stellar structure and evolution often allows for constraints on the eccentricity. If there exists a population of highly-eccentric Jupiter-sized planets, many of them will be identifiable from the light curve alone, i.e. we would deduce a large $\emin$.

\section{Generating an eccentricity posterior probability distribution}
\label{sec:mcmc}

Through an MCMC exploration---in our case implemented in the Transit Analysis Package software \citep[TAP,][]{2011G}--we can not only determine $\emin$ but impose even tighter constraints on a planet's eccentricity. For example, in \S2 we stated that a candidate whose circular density is consistent with the nominal value could have any eccentricity (i.e. for any value of eccentricity, there is an $\omega$ that satisfies $g(e,\omega) = 1$). However, for $g \sim 1$, the eccentricity posterior marginalized over $\omega$ will be dominated by low eccentricity values, even with a flat prior on the eccentricity. For example, if $e=0$, any value of $\omega$ will satisfy $g=1$, whereas only a small range of $\omega$ allow for $g = 1$ and $e > 0.9$. Thus, because we expect planetary periapses to be distributed isotropically in the galaxy, a deduced $g=1$ is most likely to truly correspond to a planet with a low eccentricity. By the same argument, the eccentricity posterior corresponding to a measured $g \ne 1$ will peak just above $\emin$. 

Of course, the transit probability also affects the eccentricity posterior distribution \citep{2008B}: an eccentric orbit with a periapse pointed towards us ($\omega=90^\circ$) is geometrically more likely to transit than a circular orbit or an eccentric orbit whose apoapse is pointed towards us. We will discuss how an MCMC exploration automatically accounts for the transit probability later in this section.

\subsection{Monte Carlo simulation of expected eccentricity and $\omega$ posteriors}
\label{subsec:monte}

To calibrate our expectations for the output of a more sophisticated MCMC parameter exploration, we first perform a Monte Carlo simulation to generate predicted posterior distributions of $e$ vs. $\omega$ via the following steps:
\begin{enumerate}
\item We begin by generating a uniform grid of $e$ and $\omega$, equivalent to assuming a uniform prior on each of these parameters. 
\item Then we calculate $g(e,\omega)$ (Equation \ref{eqn:g}) for each point $(e, \omega)$ on the grid.
\item We compute 
\begin{equation}
\label{eqn:ng}
\prob_{\rm ng} = \frac{R_\star}{a} (1-R_p/R_\star) \frac{1+e\sin\omega}{1-e^2},
\end{equation}
\noindent where $\prob_{\rm ng}$ is the probability of a non-grazing transit, for each point $(e, \omega)$ \citep[][Equation 9]{2010W}. We generate a uniform random number between 0 and 1 and discard the point if the random number is greater than the transit probability.
\item We calculate the periapse distance $\frac{a}{R_\star} (1-e)$ for each grid point and drop the point if the planet's periapse would be inside the star (effectively imposing a physically-motivated maximum eccentricity, which is most constraining for small $a/R_\star$).
\item We downsample to a subset of grid points that follows a normal distribution centered on $g$, with a width of $\sigma_g/g = 0.1$, corresponding to a 30\% uncertainty in the stellar density. To do this, we calculate the probability
\begin{equation}
\prob_g = \frac{1}{\sigma_g \sqrt{2\pi}} \exp{\left(-\frac{[g(e,\omega) - g]^2}{2 \sigma_g^2}\right)}
\end{equation}
\noindent and discard the point $(e,\omega)$ if a uniform random number is greater than $\prob_g$.
\end{enumerate}

\begin{figure*}[htbp]
\begin{centering}
\includegraphics{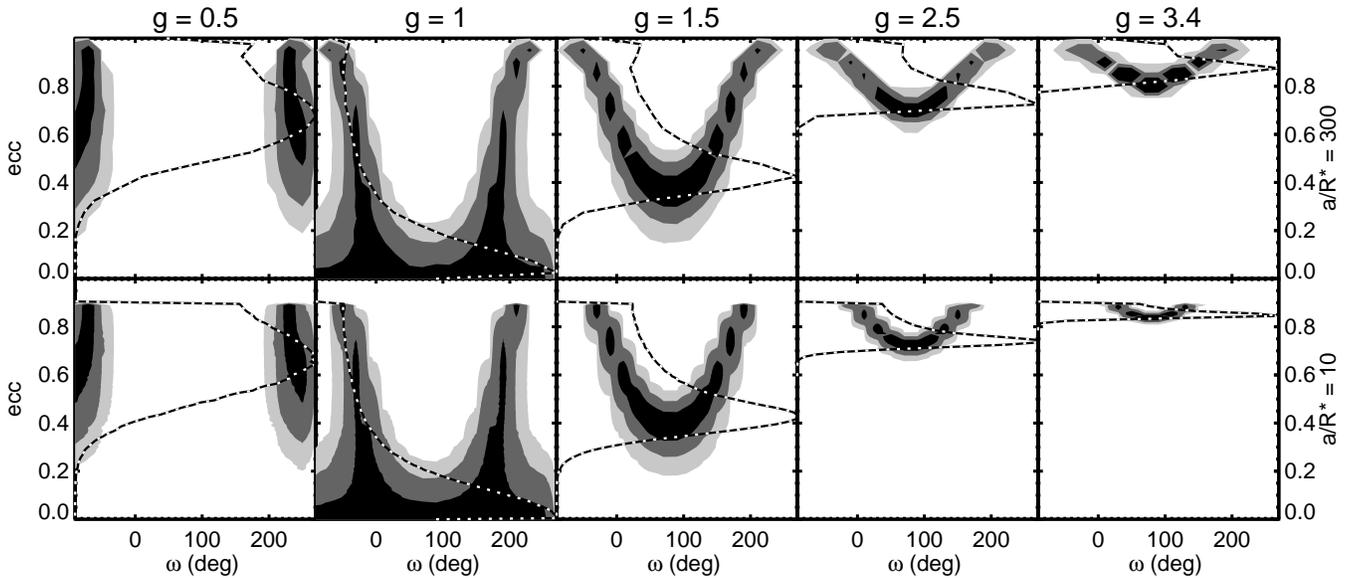}
\caption{Contoured eccentricity vs. $\omega$ posteriors from Monte Carlo simulations for representative values of $g$. The points follow a normal distribution centered at the indicated value of $g$  (columns) with a width of 10$\%$, corresponding to a 30$\%$ uncertainty in $\rho_\star$. We show the posteriors for two values of  $a/R_\star$ (rows).  The black (gray, light gray) contours represent the $\{68.3,95,99\}$\% probability density levels (i.e. 68$\%$ of the posterior is contained within the black contour). Over-plotted as a black-and-white dotted line are histograms illustrating the eccentricity posterior probability distribution marginalized over $\omega$. \label{fig:postg}}
\end{centering}
\end{figure*}

We plot the resulting posterior $e$ vs. $\omega$ distributions in Figure \ref{fig:postg} for two $a/R_\star$, one large and one small, and $R_p = 0.1$. The banana shape of the posterior results from the correlation between $e$ and $\omega$ (i.e. Equation \ref{eqn:g}).

The posteriors reveal that, rather than being inextricably entwined with $\omega$, the eccentricities deduced from $g$ are well constrained. A $\rho_{\rm circ}$ consistent with the nominal value ($g = 1$ with $\rho_\star$ constrained to within 30$\%$) is more likely to correspond to a small $e$ (e.g. the probability that $e < 0.32$ is 68.3$\%$ for $a/R_\star = 10$ and that $e < 0.35$ is 68.3$\%$ for $a/R_\star = 300$), while circular densities inconsistent with the normal values ($g$ significantly different from unity) have a well-defined minimum $e$, above which the eccentricity posterior falls off gently. For example, for $g = 2.5$ and $a/R_\star = 300$, the probability that $e > 0.69$ is 99\%. Furthermore, the eccentricity is likely to be close to this minimum eccentricity because the range of possible $\omega$ narrows as $e \rightarrow 1$. For $g = 2.5$ and $a/R_\star = 300$, the probability that $0.69 < e < 0.89$ is 95$\%$.

Next we explore how the uncertainty in $\rho_\star$ affects the eccentricity posterior, quantifying how ``loose'' this prior constraint can be. In Figure \ref{fig:postdg}, we plot eccentricity contours using a/R* = 30 for g = 1 (i.e. consistent with circular; bottom) and g = 2.5 (top) for five values of $\sigma_{\rho_\star}/\rho_\star$ assuming a normal distribution and that $\sigma_g/g = \frac{1}{3} \sigma_{\rho_\star}/\rho_\star$. For g = 2.5, the measured eccentricity is always e = 0.79; it has an uncertainty of $_{-0.06}^{+0.12 }$ for $\sigma_{\rho_\star}/\rho_\star$ = 0.01 and $_{-0.07}^{+0.12}$ for $\sigma_{\rho_\star}/\rho_\star$ = 0.5. Thus the eccentricity remains tightly constrained even for large uncertainties in the stellar density. For $g = 1$, the measured eccentricity depends more strongly on the uncertainty: $e = 0.03_{-0.03}^{+0.34}$ for $\sigma_{\rho_\star}/\rho_\star = 0.01$  and $ e = 0.24_{-0.18}^{+0.41}$ for $\sigma_{\rho_\star}/\rho_\star = 0.5$. Thus for full, ingress, and egress durations consistent with circular, a tighter constraint on the stellar density allows for a stronger upper limit on the eccentricity. However, even for a very poorly constrained $\rho_\star$, the posterior reveals that the eccentricity is most likely to be small.

\subsection{A Bayesian framework for generating posteriors}
\label{subsec:bayes}

In the Monte Carlo simulation in the previous subsection, we used random numbers to select grid points in ($e$, $\omega$) that were consistent with the light curve parameters, the prior knowledge of the stellar density, and the transit probability. An MCMC fitting routine naturally generates such a posterior in eccentricity and $\omega$ according to the following Bayesian framework.

Let the model light curve be parametrized by $e$, $\omega$, $\rho_\star$, and $X$, where $X$ represents the additional light curve parameters (i.e. orbital period, cos(inclination), radius ratio, mid transit-time, limb darkening parameters, and noise parameters). Let $D$ represent the light curve data. We wish to determine the probability of various $e$ and $\omega$ conditioned on the data, or $\prob (e,\omega,\rho_\star,X | D )$.

According to Bayes' theorem:
\begin{equation}
\prob (e,\omega,\rho_\star,X | D) \propto \prob (D | e, \omega, \rho_\star, X) \prob (e, \omega, \rho_\star, X )
\end{equation}
where the final term represents prior knowledge.

We assume a uniform prior on all the parameters except $\rho_\star$, for which we impose a prior based on the stellar parameters and their uncertainties. Therefore, we can rewrite the equation as:

\begin{equation}
\prob (e,\omega,\rho_\star,X | D) \propto \prob (D | e, \omega, \rho_\star, X) \prob (\rho_\star )
\end{equation}

Next we marginalize over $X$ and $\rho_\star$ to obtain
\begin{equation}
\prob (e,\omega| D ) \propto \int \int  \prob (D | e, \omega, \rho_\star, X) \prob (\rho_\star  ) dX d\rho_\star
\end{equation}
the two-dimensional joint posterior distribution for eccentricity and $\omega$. The first term under the integral is the likelihood of the data given $e$, $\omega$, $\rho_\star$ and $X$.  Thus a uniform prior on both these quantities naturally accounts for the transit probability because $\prob(D | e, \omega, \rho_\star, X)$ \emph{is} the transit probability; for certain values of $e$ and $\omega$, the observed transit $D$ is more likely to occur. Combinations of parameters that produce no transits are poor models, resulting in a low likelihood of the data. Evaluation of the likelihood $\prob(D | e, \omega, \rho_\star, X)$ is part of how we obtain the parameter posteriors through an MCMC exploration, the details of which we describe in the next subsection.

Finally, we can marginalize over $\omega$ to obtain
\begin{equation}
\prob (e| D) \propto \int \int \int \prob (D | e, \omega, \rho_\star, X) \prob (\rho_\star)dXd\rho_\star d\omega
\end{equation}

Thus, although stellar density, eccentricity, and $\omega$ depend degenerately on light curve properties (Equation \ref{eqn:rhoecc}), a Bayesian approach to parameter space exploration translates a loose prior on the stellar density, $\prob(\rho_\star)$, and uniform priors on the intrinsic planetary values of eccentricity and $\omega$, into a tight constraint on the planet's eccentricity. 

\subsection{Obtaining the eccentricity posterior through an MCMC sampling method}
\label{subsec:mcmcdo}

\begin{figure}[htbp]
\begin{centering}
\includegraphics{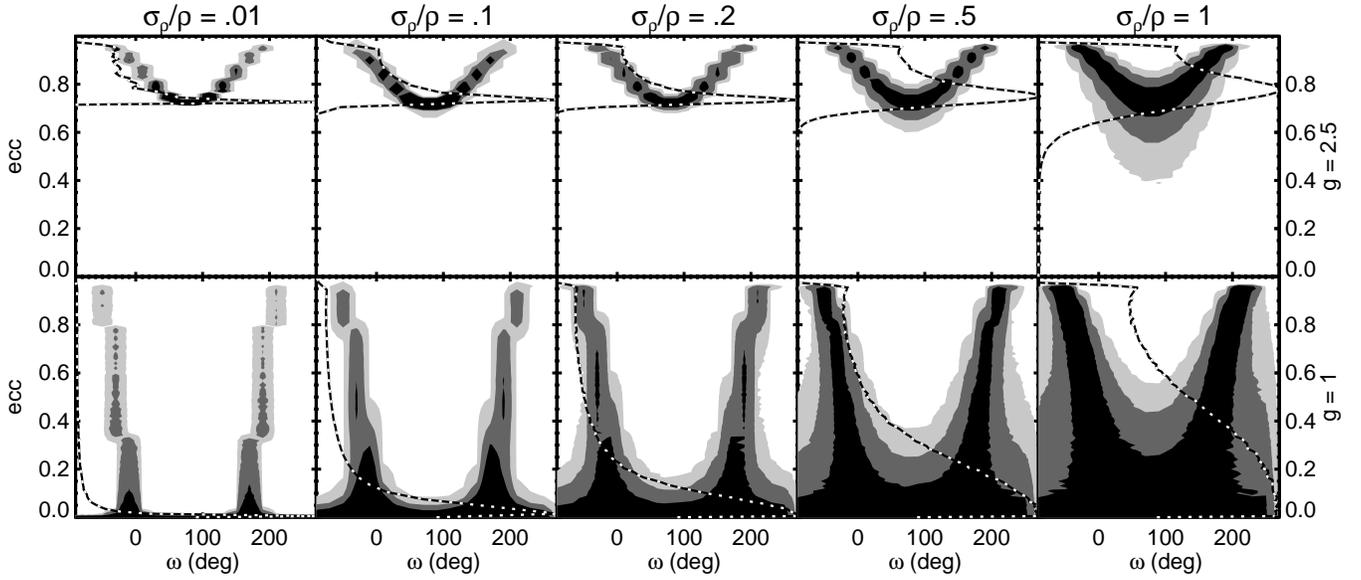}
\caption{\begin{minipage}{7.0in}Contoured eccentricity vs. $\omega$ posteriors from Monte Carlo simulations for representative values of $g$ (rows; the points follow a normal distribution centered $g$) and uncertainty in $\rho_\star$ (columns), all for $a/R_\star$ = 30.  The black (gray, light gray) contours represent the [68.3,95,99]$\%$ probability density levels. Over-plotted as a black-and-white dotted line are histograms illustrating the eccentricity posterior probability distribution marginalized over $\omega$.  \label{fig:postdg}\end{minipage}}
\end{centering}
\end{figure}

When performing light curve fits with eccentric orbital models, it is essential to use an MCMC sampling method, or some other algorithm for which the time spent in each region of parameter space is proportional to the probability. We refer the reader to \citet{2010BJ} (\S 3) for a helpful description of the MCMC method. The MCMC method can be used to minimize the $\chi^2$ (in the limit of uniform priors and Gaussian noise) or to maximize whatever likelihood function is most appropriate given one's prior knowledge. In our case, we impose a normal prior on $\rho_\star$ and account for red noise using a wavelet-based model by \citealt{2009C}. Obtaining the eccentricity posterior through an MCMC sampling method offers several advantages:

\begin{enumerate}
\item It naturally allows for marginalization over all values of $\omega$. For example, in the case of a circular density near the nominal value ($g\sim1$), the chain will naturally spend more time at low eccentricities, for which a large range of $\omega$ provide a good fit, than at high eccentricities, for which only a narrow range of $\omega$ provide a good fit.

\item It reveals and comprehensively explores complicated parameter posteriors. In particular, some of the distributions in Figure \ref{fig:postg} and \ref{fig:postdg} have banana shapes, which often cause conventional chi-squared minimization algorithms to remain stuck in the region of parameter space where they began. In contrast, an MCMC exploration will eventually fully sample the posterior distribution. \citep[See][for a pedagogical proof of this theorem.]{1995C} Because of the ``banana-shaped'' $e$ vs. $\omega$ posterior for high eccentricities (Figure \ref{fig:postg} and \ref{fig:postdg}), conventional MCMC algorithms, like TAP, require many iterations to converge and fully explore parameter space. In our case, we test for convergence by plotting $e$ and $\omega$ each as a function of chain link and assess if the exploration appears random. We also check to ensure that the $\omega$ posterior is symmetric about $\omega = 90^\circ$. Asymmetry indicates that the chains have not yet converged. We note that the variables $e\cos\omega$ and $e\sin\omega$ also have a banana-shaped posterior. When feasible, we recommend implementing an affine-invariant code such as {\tt emcee} that more efficiently explores banana-shaped posteriors \citep[e.g.][]{2012FD}. In \S\ref{subsec:gvar}, we describe how to speed up the fit convergence by using $g$ instead of $e$ as a variable while maintaining a uniform prior in $e$ and $\omega$.

\item It allows us to easily impose priors on certain parameters, such as the stellar density. If desired, one  can impose a prior on the eccentricity. In \S4, we perform an additional fit for each dataset using a Jeffrey's\footnote{We use a true Jeffrey's prior $\prob(e) \propto 1/e$, which we have not normalized because we only consider the ratio of probabilities when assessing a jump in an MCMC chain. For the fits in Section 4, for which $\emin$ is well above 0, this prior is sufficient. However, if $e = 0 $ is a possibility (i.e. for $g$ near 1), the reader may wish to use a modified Jeffrey's prior, $\prob(e) \propto 1/(e+e_0)$, where $e_0$ is the noise level. We recommend estimating an upper limit on $g$ from the uncertainty in $\rho_{\rm circ}$ and $\rho_\star$ and solving Equation (\ref{eqn:g}) for $e_0$ using $\omega = 90^\circ$.} prior on the eccentricity, which is appropriate if we wish to avoid assumptions about the magnitude of the eccentricity. Here we implement the prior through regularization (i.e. as an extra term in the jump probability).

\item It automatically accounts for the transit probability, because jumps to regions of parameter space that do not produce a transit are rejected. To address what may be a misconception, we emphasize that it is unnecessary --- and actually a double penalty --- to impose transit probability priors on the eccentricity or periapse.

\item It provides uncertainties that are more reliable than the estimates based on a simple covariance matrix (as obtained from traditional least-squares minimization) because there is no assumption that the uncertainties are normally distributed. The uncertainties fully account for complicated parameter posteriors and correlations. Therefore we can be confident in the constraints on $\rho_{\rm circ}$ even when the ingress and egress are not well-resolved. 

\end{enumerate}

We caution that although this Bayesian framework is appropriate for obtaining the posteriors of a single planet, selection effects must be carefully considered when making inferences about a population.

\subsubsection{Using $g$ as a variable for faster convergence}
\label{subsec:gvar}
Using $g$ (Equation \ref{eqn:g}) instead of $e$ as a variable in the transit fit model avoids the MCMC having to explore a banana-shaped posterior. The $g$ variable allows for faster convergence and prevents the chain from getting stuck. In order to preserve a uniform prior in $e$ and $\omega$, we must impose a prior on $g$ by adding an additional term to the likelihood function. Following the Appendix

\begin{minipage}{7.0in}
\setlength{\parindent}{1cm}
\noindent of \citet{2007BM}, the transformation from a uniform prior in $e$ to a prior in $g$ is:
\begin{eqnarray}
\prob (g ) dg &=& \prob (e) \frac{\partial e}{\partial g} dg \nonumber \\
\prob (g ) &=&  \prob (e) \frac{\partial e}{\partial g} = \frac{\sin^2\omega\left(\sin^2\omega-1\right)+g^2\left(1+\sin^2\omega\right)\pm2g\sin\omega\sqrt{\sin^2\omega-1+g^2}}{\sqrt{\sin^2\omega-1+g^2}\left(g^2+\sin^2\omega\right)^2} \nonumber \\
\end{eqnarray}
where we have assumed $\prob (e) = 1$ and for which the $+$ corresponds to $g >1 $ and the $-$ to $g<1$. .

Therefore, we add the following term to the log likelihood:

\begin{equation}
\label{eqn:glike}
\Delta \mathcal{L} = \ln \left[ \frac{\sin^2\omega\left(\sin^2\omega-1\right)+g^2\left(1+\sin^2\omega\right)\pm2g\sin\omega\sqrt{\sin^2\omega-1+g^2}}{\sqrt{\sin^2\omega-1+g^2}\left(g^2+\sin^2\omega\right)^2}\right]
\end{equation}

We demonstrate the use of this variable in \S4. We note that in our light curve fits, we use $g$ only to explore parameter space, transforming the variable to $e$ in order compute the Keplerian orbit, with no approximations, for the \citet{2002M} light curve model.

\subsection{Obtaining the eccentricity posterior from the circular-fit posterior}

The Monte Carlo exploration in \S\ref{subsec:monte} was meant to give us a handle on what the eccentricity and $\omega$ posterior should look like and how they are affected by uncertainty in $\rho_\star$. However, one could use a more formal version of this exploration to obtain posteriors of eccentricity and $\omega$ directly from the posteriors derived from circular fits to the light curve, an approach that was adopted by \citet{2012K}. One could maximize the following likelihood for the parameters $\rho_\star$, $e$, and $\omega$:

\begin{eqnarray}
\mathcal{L} = -\frac{1}{2}\frac{[ g(e,\omega)^3 \rho_\star - \rho_{\rm circ} ]^2}{\sigma_{\rho_{\rm circ}}^2} 
- \frac{1}{2} \frac{[\rho_\star- \rho_{\star,{\rm measured}}]^2}{\sigma_{\rho_{\star,{\rm measured}}}^2}
+\ln\left(\prob_{\rm ng}\right)
\end{eqnarray}

The first term in the likelihood function demands agreement with the $\rho_{\rm circ}$ derived from the circular fit to the light curve. If the $\rho_{\rm circ}$ posterior is not normal, one could replace this term with the log of the probability of $g(e,\omega)^3 \rho_\star$ given the $\rho_{\rm circ}$ posterior. Note that $g(e,\omega)$ can either be computed from the approximation in Equation (\ref{eqn:g}) or by solving and integrating Kepler's equation to obtain the mean ratio of the transiting planet's velocity to its  Keplerian velocity over the course of the transit. The second term is the prior on $\rho_\star$ from the stellar parameters independently measured from spectroscopy (or asteroseismology). The final term is the probability of a non-grazing transit (Equation \ref{eqn:ng}). If one uses the variable $g$ instead $e$, one should add Equation (\ref{eqn:glike}) to the likelihood. We warn that this likelihood function drops constants, so although it can be used to generate parameter posteriors, it should not be used 
\end{minipage}
\clearpage
\noindent 
to compute the Bayesian evidence quantity.

In the next section, we demonstrate that this approach yields the same eccentricity and $\omega$ posteriors as directly fitting for the eccentricity from the light curve.
\section{Demonstration: measuring the eccentricities of transiting Jupiters}

To demonstrate that the duration aspect of the photoeccentric effect allows for precise and accurate measurements of a transiting planet's eccentricity from the light curve alone, we apply the method described in \S\ref{sec:mcmc} to several test cases. In \S\ref{subsec:lit} we measure the eccentricity of a transiting planet that has a known eccentricity from RV measurements. In \S\ref{subsec:inject} we inject a transit into short and long cadence \kep\ data and compare the resulting e and $\omega$ posteriors. In \S\ref{subsec:koi}, we measure the eccentricity of a \kep candidate that has only long-cadence data available.

\subsection{HD~17156\,b: a planet with a large eccentricity measured from RVs}
\label{subsec:lit}

HD~17156\,b was discovered by the Next 2000 Stars (N2K) Doppler survey \citep{2005F,2007F}. \citet{2007F} reported that the planet has a large orbital eccentricity of $e = 0.67 \pm 0.08$. We identified this planet and the relevant references using {\tt exoplanets.org} \citep{2011WF}. \citet{2007BA} reported several partial transits observed by small-telescope observers throughout the Northern Hemisphere, and \citet{2009B} and \citet{2009W} observed full transits using high-precision, ground-based photometry. Here we demonstrate that the planet's eccentricity could have been measured from the transit light curve data alone.

\begin{figure}[htbp]
\begin{centering}
\includegraphics{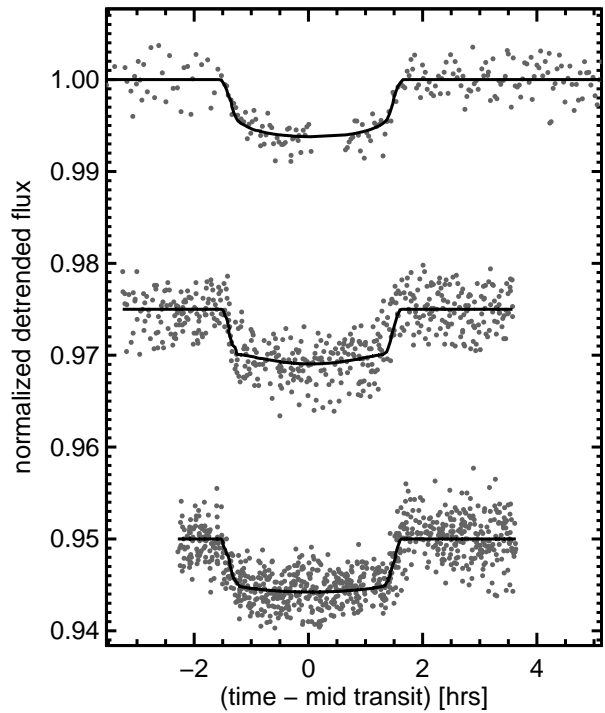}
\caption{Light curves of HD~17156 from \citet{2009B} (top) and \citet{2009W} (middle, bottom). A set of eccentric model light curves drawn from the posterior are plotted as solid lines.\label{fig:lc}}
\end{centering}
\end{figure}

We simultaneously fit three light curves (Figure \ref{fig:lc}), one from \citet{2009B} and two from \citet{2009W} using TAP \citep{2011G}, which employs an MCMC technique to generate a posterior for each parameter of the Mandel and Agol (2002) transit model. Time-correlated, ``red'' noise is accounted for using the \citet{2009C} wavelet-based likelihood function. To achieve the $2^N$ (where N is an integer) data points required by the wavelet-based likelihood function without excessive zero-padding, we trimmed the first \citet{2009W} light curve from 523 data points to 512 data points by removing the last 11 data points in the time series. Initially, we fixed the candidate's eccentricity at 0 and fit for $\rho_{\rm circ}$ with no prior imposed, to see how much it differs from the well-measured value of $\rho_\star$. Then we refitted the transit light curves with a normal prior imposed on the stellar density, this time allowing the eccentricity to vary. In both cases, we treated the limb darkening coefficients following the literature: we fixed the coefficients for the \citet{2009B} light curve and left the coefficients free for the \citet{2009W} light curves. Following \citet{2009W}, we also included linear extinction free parameters for the two \citet{2009W} light curves. (The published \citealt{2009B} light curve was already pre-corrected for extinction.)

\begin{figure}[htbp]
\begin{centering}
\includegraphics{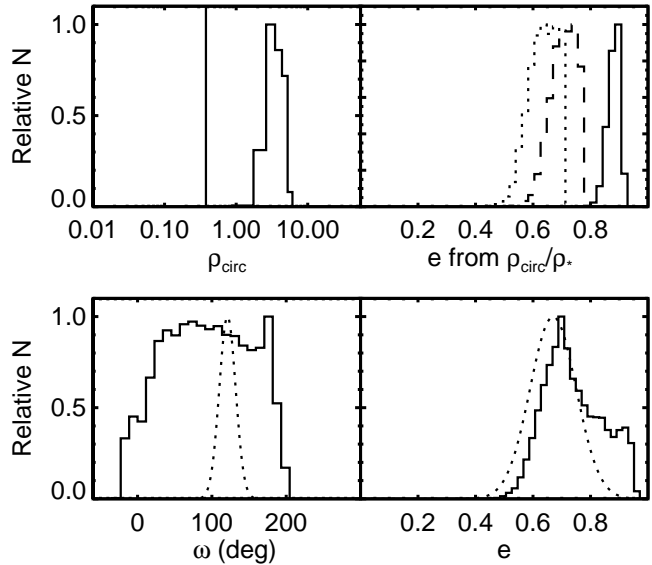}
\caption{Posterior distributions of $e$ and $\omega$ for the HD~17156 transiting system, with eccentricity fixed at  0 (row 1) and free to vary (row 2). Row 1: Left: $\rho_\star$ derived from circular fit. The solid line marks the nominal value. Right: Posterior distribution for eccentricity solving Equation (\ref{eqn:arecc}) for $\omega=0$ (solid line), $\omega=45^\circ$ (dashed line), and $\omega=90^\circ$ (dotted line). Row 2: Left: Posterior distribution for $\omega$ from eccentric fit (i.e. a fit to the light curve in which the eccentricity is a free parameter; solid). Gaussian illustrating posterior from \citet{2007F} RV fit (dotted line). Right: Same for eccentricity posterior.\label{fig:lit}}
\end{centering}
\end{figure}

\begin{figure}[htbp]
\begin{centering}
\includegraphics{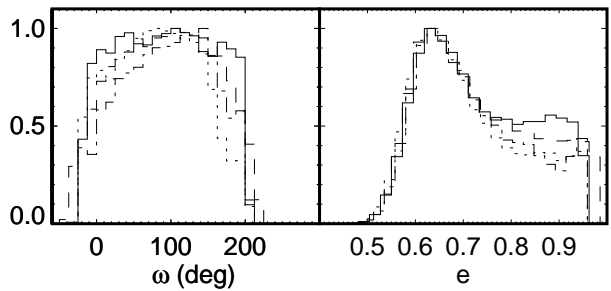}
\caption{Left: Posterior distribution for $\omega$ for a fit to the light curve using $g$ as a free parameter with a uniform prior on the eccentricity (sold line) and Jeffrey's prior (dotted line). Posterior distribution using $e$ instead of $\omega$ as a free parameter (dot-dashed line). Posterior distribution using method described in \S3.4 (dashed line). Right: Same as left, for eccentricity posterior. \label{fig:lit3}}
\end{centering}
\end{figure}

Figure \ref{fig:lit} shows posterior distributions from a circular fit (top row) and an eccentric fit (bottom row) with a prior imposed on the stellar density from \citet{2011GM}. In Figure \ref{fig:lit3}, we compare the posteriors generated from a) the eccentric fit to the light curve using $g$ as a parameter (with a prior imposed to maintain a uniform eccentricity prior; Equation (\ref{eqn:glike}) to posteriors generated using: b) a Jeffrey's prior on the eccentricity, c) $e$ instead of $g$ as a free parameter (to demonstrate that they are equivalent), and d) the likelihood-maximization method described in \S3.4, using the posterior of $\rho_{\rm circ}$ from the circular fit. The four sets of posteriors closely resemble one another. The computation times were about 1 day for the circular fit, about 1 day for the eccentric fit using $g$ as a parameter, several days for the eccentric fit using $e$ as a parameter, and thirty minutes for the likelihood maximization method of \S3.4. Note that the final method requires the best-fitting parameters resulting from a circular fit to the light curve, including accurate parameter posteriors. We therefore caution against using the parameters listed in the \kep public data releases for this purpose because those values are the result of a least-squares fit and make the assumption of normally distributed parameter uncertainties. However, if one has already precomputed circular fits using an MCMC algorithm that incorporates red noise and limb darkening---as we have done for all of the Jupiter-sized KOIs (\S 5)---the final method (\S 3.4) is advantageous because of the decreased computation time.

Based on the circular fit alone, we would infer $g(\emin,\pi/2)$ = 2.0, corresponding to a minimum eccentricity of $\emin = 0.61$. From the eccentric fit, we obtain a value of $e = 0.71_{-0.09}^{+0.16}$ using a uniform prior on the eccentricity and $e = 0.69_{-0.09}^{+0.16}$ using a Jeffrey's prior. Therefore, we could have deduced the eccentricity determined from 33 RV measurements --- $ e = 0.67 \pm 0.08$ \citep{2007F} --- from these three transit light curves alone.

\begin{figure}[htbp]
\begin{centering}
\includegraphics{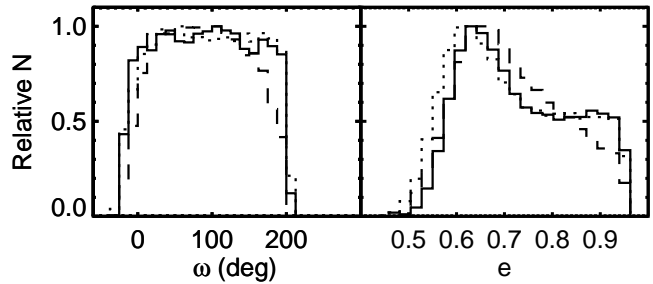}
\caption{Posterior distributions of $e$ and $\omega$ for the HD~17156 transiting system, with three different priors on the stellar density: the density measured by \citet{2011GM} (solid); the density measured by \citet{2011GM} with uncertainties enlarged to $\sigma_{\rho_\star}/\rho_\star$ = 0.2, (dashed) and the density based on the stellar parameters from \citet{2009W} (dotted).\label{fig:litcomp}}
\end{centering}
\end{figure}

\begin{figure*}[htbp]
\begin{centering}
\includegraphics{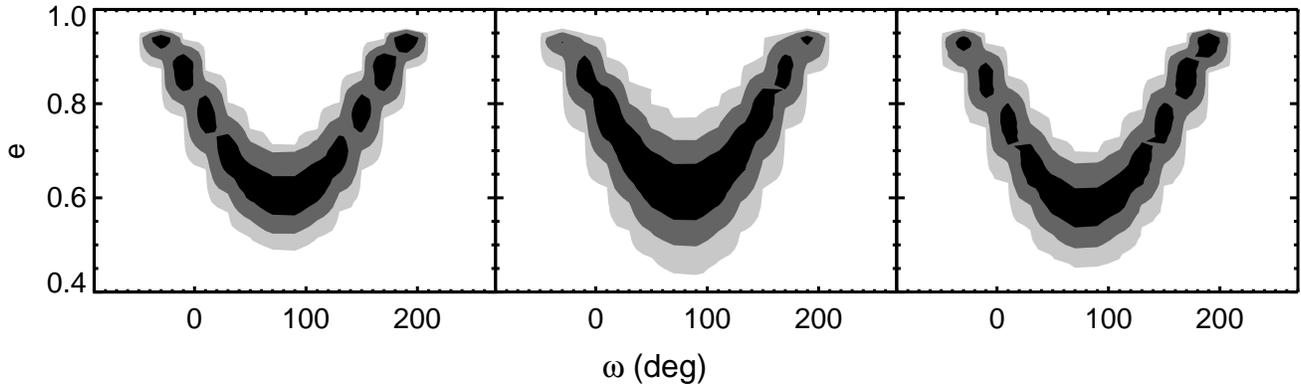}
\caption{Eccentricity vs. $\omega$ posterior distributions  for HD~17156~b based on fits using a prior on the stellar density from \citet{2011GM} (left); \citet{2011GM} with error bars enlarged to 20$\%$ (middle); and \citet{2009W} (left). \label{fig:eom}}
\end{centering}
\end{figure*}

The host star has a particularly well-constrained density from asteroseismology \citep{2011GM}. We artificially enlarge the error bars on the stellar density from 1$\%$ to 20$\%$ and repeat the fitting procedure, obtaining an eccentricity of $e = 0.70_{-0.09}^{+0.14}$. We also repeat the fitting procedure with a density derived from the stellar parameters $M_\star$ and $R_\star$ determined by \citet{2009W} from isocrone fitting. This ``pre-asteroseismology'' density has an uncertainty of 10$\%$ and, moreover, is about 5$\%$ larger than the value measured by \citet{2011GM}. We obtain an eccentricity of $ e = 0.70_{-0.11}^{+0.16}$. In Figure \ref{fig:litcomp} and \ref{fig:eom}, we plot the resulting posterior distributions, which are very similar. Therefore, even with uncertainties and systematics in the stellar density, we can measure a transiting planet's eccentricity to high precision and accuracy.

\subsection{Short vs. long cadence \kep data}
\label{subsec:inject}

\citet{2010Kb} explored in detail the effects of long integration times and binning on transit light curve measurements, with a particular focus on long-cadence \kep data. He demonstrated that by binning a finely-sampled model to match the cadence of the data, as TAP has implemented, one can fit accurate (though less precise than from short cadence data) light curve parameters. Using short and long cadence \kep data of a planet with known parameters (TrES-2-b), he validated this approach.

Here we explore, through a test scenario of an eccentric planet injected into short and long \kep data, whether this approach holds (as one would expect) for fitting an eccentric orbit and what value short-cadence data adds to the constraint on eccentricity. We chose parameters for the planet typical of an eccentric Jupiter and main-sequence host star: $P = 60$ days, $i = 89.5^\circ$, $R_p/R_\star=$ 0.1, $e = $ 0.8, $\omega = 90^\circ$, $M_\star = R_\star = 1$, and limb darkening parameters $\mu_1 = \mu_2 = 0.3$. We considered the situation in which long cadence data is available for Q0-Q6 but short-cadence is available only for one quarter (or may be in the future). We retrieved Q0-Q6 data from the Multimission Archive at the Space Telescope Science Institute (MAST)  for \kep target star KIC 2306756, selected because it has both long and short cadence data. Then we applied the TAP MCMC fitting routine to fit a) one short-cadence transit (fixing the period at 60 days) that took place in a single segment of short-cadence data and b) all seven long-cadence transits.

\begin{figure*}[htbp]
\begin{centering}
\includegraphics{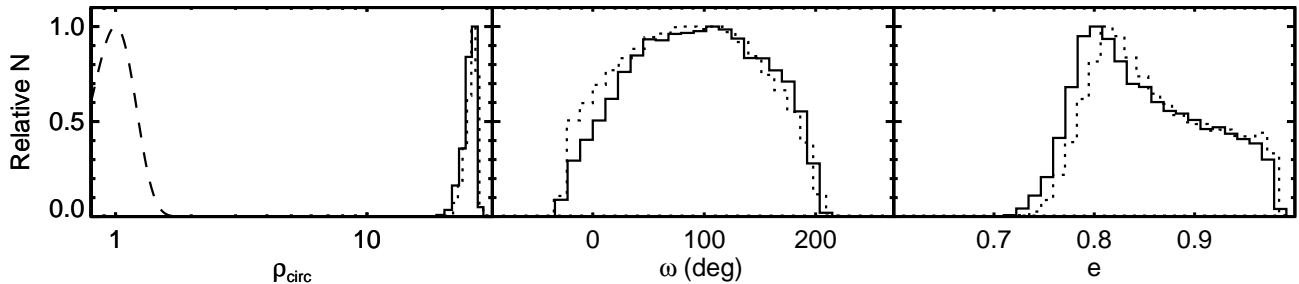}
\caption{Posterior distributions of $e$ and $\omega$ for an injected, artificial transit, with eccentricity fixed at  0 (panel 1) and free to vary (panel 2-3). The sold curves are from a fit to seven light curves from the long-cadence data and the dotted to a single light curve from the short cadence data. Left: $\rho_\star$ derived from circular fit. The dashed curve represents the nominal value and its uncertainty. Middle: Posterior distribution for $\omega$ from eccentric fit (solid line). Right: Eccentricity posterior. \label{fig:inject}}
\end{centering}
\end{figure*}

As in \S4.1, we performed one set of fits fixing the orbit as circular and another set with $g$ and $\omega$ as free parameters, imposing a prior on the stellar density corresponding to a 20$\%$ uncertainty in the stellar density and a prior on $g$ from a uniform prior in $e$ and $\omega$ (Equation \ref{eqn:glike}). In both cases, we allowed the limb darkening to be a free parameter. We plot the resulting posterior distributions of eccentricity and $\omega$ in Figure~\ref{fig:inject}. From the circular fits, the constraint on $\rho_{\rm circ}$ is somewhat stronger from the short cadence data ($26.3^{+1.0}_{-1.6}~\rho_\odot$) than from the long cadence data $(25.9^{+1.0}_{-2.7}~\rho_\odot$), as \citet{2010Kb} found. From the short cadence data, we measure an eccentricity of $e = 0.85_{-0.05}^{+0.08}$ with a uniform prior on the eccentricity and $e = 0.85_{-0.05}^{+0.07}$ with a Jeffrey's prior. From the long cadence data, we measure an eccentricity of $e = 0.84_{-0.05}^{+ 0.08}$ with a uniform prior on the eccentricity and $e = 0.84_{-0.04}^{+ 0.07}$ with a Jeffrey's prior. Therefore, the long cadence data is sufficient to obtain a precise eccentricity measurement. In this case, the 20$\%$ uncertainty in the stellar density dominated over the constraint from the transit light curve on $\rho_{\rm circ}$; however, for very well-constrained stellar properties, we would expect the greater precision of the short cadence data to allow for a tighter constraint on the eccentricity (see Figure \ref{fig:postdg}).

\subsection{KOI~686.01, a moderately eccentric, Jupiter-sized \kep candidate}
\label{subsec:koi}

\begin{figure}[htbp]
\begin{centering}
\includegraphics{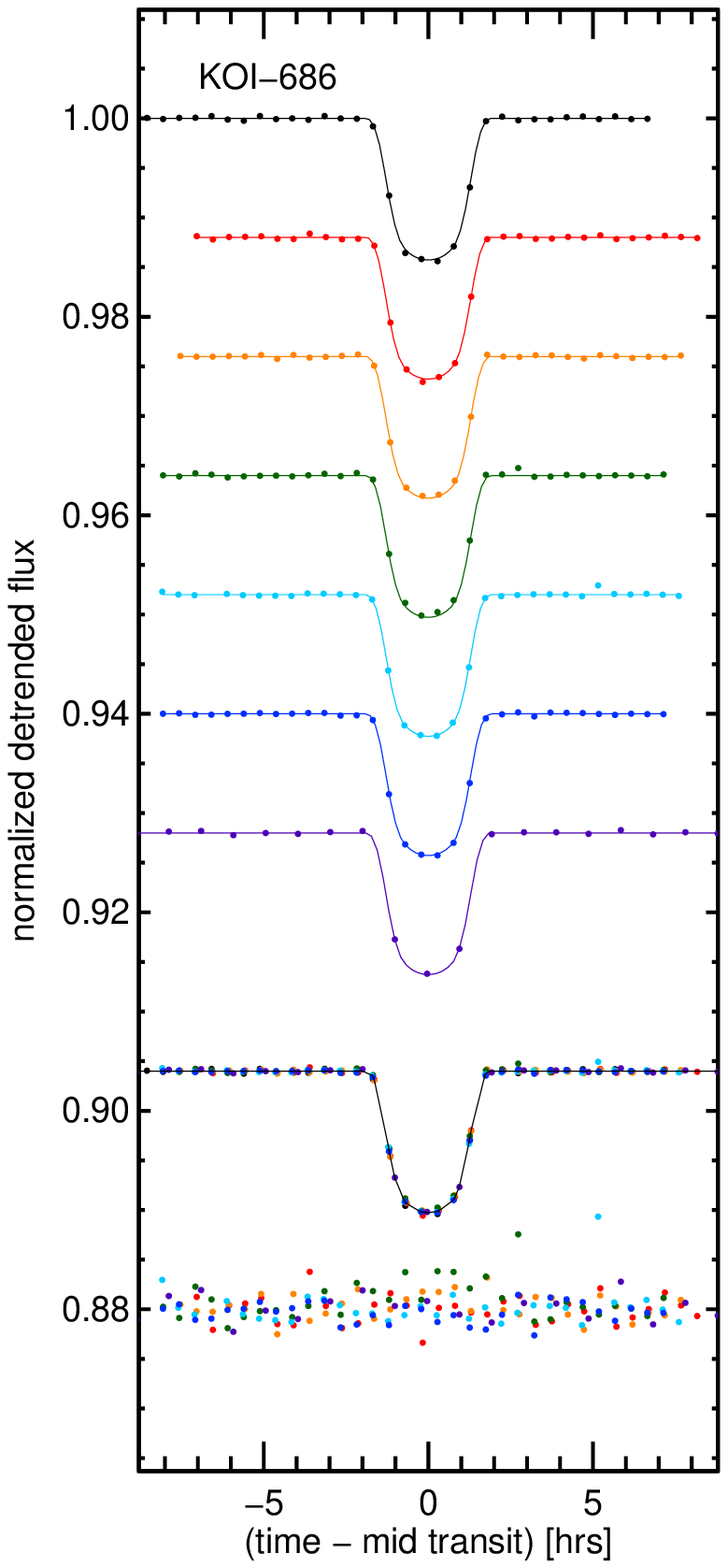}
\caption{Light curves of KOI~686. A set of eccentric model light curves drawn from the posterior are plotted as solid lines. The second-from-bottom curve  is a compilation of all the light curves. The bottom points are the residuals multiplied by 10. \label{fig:lc_koi}}
\end{centering}
\end{figure}

KOI~686.01 was identified by \citet{2011BKB} and \citet{2012B} as a 11.1 $R_{\rm Earth}$ candidate that transits its host star every 52.5135651 days. We retrieved the Q0-Q6 data from MAST and detrended the light curve using {\tt AutoKep} \citep{2011G}. We plot the light curves in Figure \ref{fig:lc_koi}.

We obtained a spectrum of KOI~ 686 using the HIgh Resolution Echelle Spectrometer (HIRES) on the Keck I Telescope \citep{1994V}. The spectrum was obtained with the red cross-disperser and $0\farcs86$ slit using the standard setup of the California Planet Survey (CPS), but with the iodine cell out of the light path. The extracted spectrum has a median signal-to-noise ratio of 40 at 5500~$\AA$, and a resolution $\lambda/\Delta\lambda \approx 55,000$. To estimate the stellar temperature, surface gravity, and metallicity, we use the {\tt SpecMatch} code, which searches through the CPS's vast library of stellar spectra for stars with \emph{Spectroscopy Made Easy} \citep[SME;][]{1996V,2005V} parameters and finds the best matches. The final values are the weighted mean of the 10 best matches. We then interpolate these stellar parameters onto the Padova stellar evolution tracks to obtain a stellar mass and radius. We checked these values using the empirical relationships of \citet{2010T}. We find $\rho_\star = 1.02^{+0.45}_{-0.29} $~$\rho_\odot$ (the other stellar parameters for this KOI and parameters for other KOI will be published as part of another work, Johnson et al. 2012, in prep). 

\begin{figure}[htbp]
\begin{centering}
\includegraphics{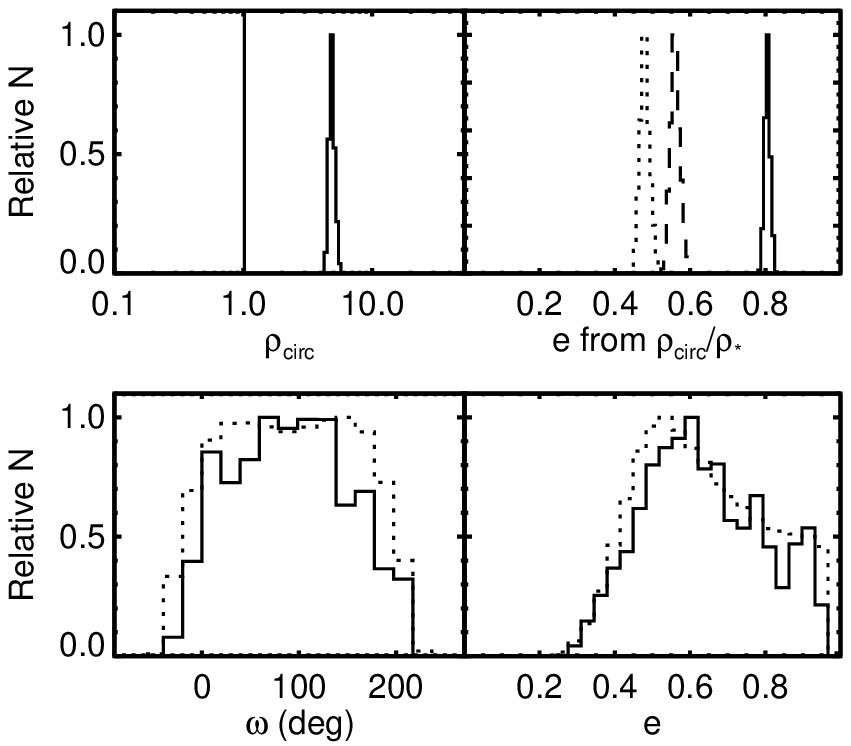}
\caption{Posterior distributions for KOI 686.01 with eccentricity fixed at  0 (row 1) and free to vary (row 2). Row 1: Left: $\rho_\star$ derived from circular fit. The solid line marks the nominal value. Right: Posterior distribution for eccentricity solving Equation (\ref{eqn:arecc}) for $\omega=0$ (solid line), $\omega=45^\circ$ (dashed line), and $\omega=90^\circ$ (dotted line). Row 2: Left: Posterior distribution for $\omega$ from eccentric fit (solid). Posterior distribution using method from \S3.4 (dotted). Right: Same as left, for eccentricity posterior. \label{fig:koi}}
\end{centering}
\end{figure}

We then fit circular and eccentric orbits to the transit light curve, as described above, binning the model light curves to match the 30-minute cadence of the data. We impose a normal prior on the limb-darkening coefficients based on the values from \citet{2010S}. Figure \ref{fig:koi} shows posterior distributions from a circular fit (top row) and an eccentric fit (bottom row) with a prior imposed on the stellar density. We measure the eccentricity to be $e = 0.62_{-0.14}^{+0.18}$.

We caution that this candidate has not yet been validated; \citet{2011MJ} estimate a false-positive probability of $8\%$. If the candidate is a false positive, its orbit (and other properties, such as its radius) is likely to be different from that inferred. However, we note that if the candidate is a background binary or hierarchical triple and is actually larger than a planet, the inferred eccentricity would actually be higher (i.e. if the candidate is actually larger, it must be moving through its ingress and egress even faster), unless KOI 686 is not the primary and the primary has a higher density than KOI 686. Another possibility, if the candidate is false positive, is that the assumption of $M_p \ll M_\star$ may no longer hold and $\rho_{\circ}$ (Equation \ref{eqn:rhoecc}) should be compared to $\rho_\star+\rho_{\rm companion}$ rather than $\rho_star$ to obtain $g$. However, even if $\rho_{\rm companion} \sim \rho_\star$, the error in $g$ would be only $(\frac{1}{2})^3 = 12.5\%$.

\citet{2012SD} recently found a false positive rate of $35\%$ for Jupiter-sized candidates, comprised of brown dwarfs, undiluted eclipsing binaries, and diluted eclipsing binaries. In the case of diluted eclipsing binaries, the blend effects that we discussed in \S2 could be larger than we considered. However, \citet{2012M} notes that most of the false positives that \citet{2012SD} discovered through radial-velocity follow-up already exhibited V-shapes or faint secondary eclipses in their light curves. In the search for highly eccentric Jupiters, we recommend a careful inspection of the transit light curve for false-positive signatures and, when possible, a single spectroscopic observation and adaptive-optics imaging to rule out false-positive scenarios. 

If the planetary nature of this object is confirmed, it will be one of a number of Jupiter-sized planets with orbital periods of 10-100 days and moderate eccentricities, but the first in the \kep\ sample with a photometrically-measured eccentricity. Many previously known, moderately-eccentric planets have orbits inside the snow line; their eccentricities are thought to be signatures of the dynamical process(es) that displaced them from their region of formation.

\section{A plan for distilling highly-eccentric Jupiters from the \kep sample}
\label{sec:distill}

To test the HEM hypothesis (S12), we are ``distilling'' highly-eccentric, Jupiter-sized planets --- proto hot Jupiters --- from the sample of announced \kep candidates using the publicly released \kep light curves \citep{2011BKB,2012B}. To identify planets that must be highly eccentric, we are refitting the \kep light curves of all the Jupiter-sized candidates using the TAP. Initially, we fix the candidate's eccentricity at 0. We identify candidates whose posteriors for $\rho_{\rm circ}$ are wildly different than the nominal value $\rho_\star$ from the KIC. From this subset of objects, we obtain spectra of the host stars. We refine the stellar parameters using {\tt SpecMatch}, interpolate them onto the Padova stellar evolution tracks to obtain a stellar mass and radius, and check the inferred $M_\star$ and $R_\star$ using the empirical relationships of \citet{2010T}. We validate the candidate using the method outlined in \citet{2012M}. Finally, we refit the transit light curves with a prior imposed on the stellar density, this time allowing the eccentricity to vary. This process will allow to us easily identify the most unambiguous highly-eccentric hot Jupiters.

\section{Discussion}
\label{sec:discuss}

Measuring a transiting planet's orbital eccentricity was once solely the province of radial-velocity observations. Short-period planets were discovered by transits and followed-up with RVs, which sometimes revealed a sizable eccentricity (e.g. HAT-P-2b, \citealt{2007BK}; CoRoT-10b, \citealt{2010BS}). Long-period planets---which, based on the RV distribution, are more commonly eccentric---were discovered by radial-velocity measurements and, on lucky occasions, found to transit (e.g. HD~17156\,b, \citealt{2007F}, the planet discussed in \S\ref{subsec:lit}, as well as HD~806066\,b, \citealt{2001N}). But now, from its huge, relatively unbiased target sample size of 150,000 stars, \kep has discovered a number of long-period, transiting candidates. Among these are likely to be a substantial number of eccentric planets (S12), which have enhanced transit probabilities \citep{2012KC}. \citet{2011M,2012KC} and \citet{2012P} have characterized the eccentricity distributions of these candidates based on \kep photometry. \citet{2012K} are employing MAP to measure the eccentricities of planets in systems in which multiple planets transits. Here we have demonstrated that it is also possible to constrain an individual planet's eccentricity from a set of high signal-to-noise transits using a Bayesian formalism that employs relatively loosely-constrained priors on the stellar density. The technique we have presented can be applied to any transit light curve, as we did in \S\ref{subsec:lit}, for HD~17156\,b using ground-based photometry. Comparing this technique to \citet{2012K}'s MAP, MAP is more model independent -- requiring no knowledge at all of the stellar density -- but our technique is applicable to single transiting planets, as Jupiter-sized \kep candidates tend to be \citep[e.g.][]{2011LR}. We are the process of fitting the orbits of all Jupiter-sized \kep candidates, which will lead to the following prospects:

\begin{enumerate}
\item For candidates with host stars too faint for RV follow-up (65$\%$ of candidates in \citealt{2011BKB} are fainter than \kep magnitude 14), our technique will provide an estimate of the planet's eccentricity. We may also be able to deduce the presence of companions from transit timing variations, thereby allowing us to search for ``smoking gun'' perturbers that may be responsible for the inner planet's orbital configuration. In a companion paper (Dawson et al. 2012, in prep), we present the validation and characterization of a KOI with a high, photometrically-measured eccentricity and transit timing variations.

\item For candidates bright enough for follow-up RV measurements, the eccentricity and $\omega$ posteriors from photometric fits allow us to make just a few optimally timed radial velocity measurements to pinpoint the planet's eccentricity, the mass and host-star density, instead of needing to devote precious telescope time to sampling the full orbital period. The tight constraints on eccentricity from photometry alone can be combined with radial-velocity measurements to constrain the candidate's orbit---either by fitting both datasets simultaneously or by using the posteriors from the photometry as priors for fitting a model to the RVs. To maximize the information gain, the prior on the stellar density should remain in place. This serves as an additional motivation for measuring the spectroscopic properties of candidate host stars in the \kep\ field.

\item We can also measure the spin-orbit angles of the candidates orbiting the brightest stars with Rossiter-McLaughlin measurements. Then we can compare the distribution of spin-orbit angles of those planets we have identified as eccentric with the distribution of those we have constrained to be most likely circular.

\item S12 argue that HEM mechanisms for producing hot Jupiters should also produce a population of highly eccentric ($e > 0.9$) proto hot Jupiters and predict that we should find 3-5 in the \kep sample. Moreover, \emph{Kepler}'s continuous coverage may offer the best prospect for detecting highly eccentric planets, against which RV surveys are biased \citep{2006J,2009O}. In \S\ref{sec:distill}, we described our process for distilling highly-eccentric Jupiters from the \kep sample.

\end{enumerate}

The \kep sample has already revealed a wealth of information about the dynamics and architectures of planetary systems \citep[e.g.][]{2011L,2012FL} but primarily for closely-packed systems of low mass, multiple-transiting planets. Measuring the eccentricities of individual, Jupiter-sized planets in the \kep will allow us to investigate a different regime: planetary systems made up of massive planets that potentially underwent violent, mutual gravitational interactions followed by tidal interactions with the host star.

\acknowledgments We are thankful for the helpful and positive feedback from the anonymous referee. R.I.D. gratefully acknowledges support by the National Science Foundation Graduate Research Fellowship under grant DGE-1144152, clear and constant guidance from chapter \citet{2010W}, and the ministry and fellowship of the Bayesian Book Club. J.A.J. thanks Avi Loeb and the ITC for hosting him as part of their visitors program, thereby allowing the authors to work together in close proximity at the CfA during the completion of this work. We thank Sarah Ballard, Zachory Berta, Joshua Carter, Courtney Dressing, Subo Dong, Daniel Fabrycky, Jonathan Irwin, Boaz Katz, David Kipping, Timothy Morton, Norman Murray, Ruth Murray-Clay, Peter Plavchan, Gregory Snyder, Aristotle Socrates, and Joshua Winn for helpful discussions. Several colleagues provided helpful and inspiring comments on a manuscript draft: Joshua Carter (who, in addition to other helpful comments, suggested the procedure described in \S3.4), Daniel Fabrycky, Eric Ford, David Kipping, and Ruth Murray-Clay. Special thanks to J. Zachary Gazak for helpful modifications to the TAP code. 

This paper includes data collected by the \kep mission. Funding for the \kep mission is provided by the NASA Science Mission directorate. Some of the data presented in this paper were obtained from the Multimission Archive at the Space Telescope Science Institute (MAST). STScI is operated by the Association of Universities for Research in Astronomy, Inc., under NASA contract NAS5-26555. Support for MAST for non-HST data is provided by the NASA Office of Space Science via grant NNX09AF08G and by other grants and contracts.

This research has made use of the Exoplanet Orbit Database and the Exoplanet Data Explorer at {\tt exoplanets.org}.

\bibliography{./PEHJ} \bibliographystyle{apj} 

\begin{thebibliography}{57}
\expandafter\ifx\csname natexlab\endcsname\relax\def\natexlab#1{#1}\fi

\bibitem[{{Alibert} {et~al.}(2005){Alibert}, {Mordasini}, {Benz}, \&
  {Winisdoerffer}}]{2005A}
{Alibert}, Y., {Mordasini}, C., {Benz}, W., \& {Winisdoerffer}, C. 2005, \aap,
  434, 343

\bibitem[{{Bakos} {et~al.}(2007){Bakos}, {Kov{\'a}cs}, {Torres}, {Fischer},
  {Latham}, {Noyes}, {Sasselov}, {Mazeh}, {Shporer}, {Butler}, {Stefanik},
  {Fern{\'a}ndez}, {Sozzetti}, {P{\'a}l}, {Johnson}, {Marcy}, {Winn}, {Sip{\H
  o}cz}, {L{\'a}z{\'a}r}, {Papp}, \& {S{\'a}ri}}]{2007BK}
{Bakos}, G.~{\'A}., {Kov{\'a}cs}, G., {Torres}, G., {Fischer}, D.~A., et al. 2007, \apj, 670, 826

\bibitem[{{Barbieri} {et~al.}(2009){Barbieri}, {Alonso}, {Desidera},
  {Sozzetti}, {Martinez Fiorenzano}, {Almenara}, {Cecconi}, {Claudi},
  {Charbonneau}, {Endl}, {Granata}, {Gratton}, {Laughlin}, {Loeillet}, \&
  {Amateur Consortium}}]{2009B}
{Barbieri}, M., {Alonso}, R., {Desidera}, S., {Sozzetti}, A., et al. 2009, \aap, 503, 601

\bibitem[{{Barbieri} {et~al.}(2007){Barbieri}, {Alonso}, {Laughlin},
  {Almenara}, {Bissinger}, {Davies}, {Gasparri}, {Guido}, {Lopresti},
  {Manzini}, \& {Sostero}}]{2007BA}
{Barbieri}, M., {Alonso}, R., {Laughlin}, G., {Almenara}, J.~M., et al. 2007, \aap, 476, L13

\bibitem[{{Barnes}(2007)}]{2007B}
{Barnes}, J.~W. 2007, \pasp, 119, 986

\bibitem[{{Batalha} {et~al.}(2010){Batalha}, {Borucki}, {Koch}, {Bryson},
  {Haas}, {Brown}, {Caldwell}, {Hall}, {Gilliland}, {Latham}, {Meibom}, \&
  {Monet}}]{2010B}
{Batalha}, N.~M., {Borucki}, W.~J., {Koch}, D.~G., {Bryson}, S.~T., et al. 2010, \apjl, 713, L109

\bibitem[{{Batalha} {et~al.}(2012){Batalha}, {Rowe}, {Bryson}, {Barclay},
  {Burke}, {Caldwell}, {Christiansen}, {Mullally}, {Thompson}, {Brown},
  {Dupree}, {Fabrycky}, {Ford}, {Fortney}, {Gilliland}, {Isaacson}, {Latham},
  {Marcy}, {Quinn}, {Ragozzine}, {Shporer}, {Borucki}, {Ciardi}, {Gautier},
  {Haas}, {Jenkins}, {Koch}, {Lissauer}, {Rapin}, {Basri}, {Boss}, {Buchhave},
  {Charbonneau}, {Christensen-Dalsgaard}, {Clarke}, {Cochran}, {Demory},
  {Devore}, {Esquerdo}, {Everett}, {Fressin}, {Geary}, {Girouard}, {Gould},
  {Hall}, {Holman}, {Howard}, {Howell}, {Ibrahim}, {Kinemuchi}, {Kjeldsen},
  {Klaus}, {Li}, {Lucas}, {Morris}, {Prsa}, {Quintana}, {Sanderfer},
  {Sasselov}, {Seader}, {Smith}, {Steffen}, {Still}, {Stumpe}, {Tarter},
  {Tenenbaum}, {Torres}, {Twicken}, {Uddin}, {Van Cleve}, {Walkowicz}, \&
  {Welsh}}]{2012B}
{Batalha}, N.~M., {Rowe}, J.~F., {Bryson}, S.~T., {Barclay}, T., et al. 2012, arXiv:1202.5852

\bibitem[{{Bonomo} {et~al.}(2010){Bonomo}, {Santerne}, {Alonso}, {Gazzano},
  {Havel}, {Aigrain}, {Auvergne}, {Baglin}, {Barbieri}, {Barge}, {Benz},
  {Bord{\'e}}, {Bouchy}, {Bruntt}, {Cabrera}, {Collier Cameron}, {Carone},
  {Carpano}, {Csizmadia}, {Deleuil}, {Deeg}, {Dvorak}, {Erikson},
  {Ferraz-Mello}, {Fridlund}, {Gandolfi}, {Gillon}, {Guenther}, {Guillot},
  {Hatzes}, {H{\'e}brard}, {Jorda}, {Lammer}, {Lanza}, {L{\'e}ger}, {Llebaria},
  {Mayor}, {Mazeh}, {Moutou}, {Ollivier}, {P{\"a}tzold}, {Pepe}, {Queloz},
  {Rauer}, {Rouan}, {Samuel}, {Schneider}, {Tingley}, {Udry}, \&
  {Wuchterl}}]{2010BS}
{Bonomo}, A.~S., {Santerne}, A., {Alonso}, R., {Gazzano}, et al. 2010, \aap, 520, A65

\bibitem[{{Borucki} {et~al.}(2011){Borucki}, {Koch}, {Basri}, {Batalha},
  {Brown}, {Bryson}, {Caldwell}, {Christensen-Dalsgaard}, {Cochran}, {DeVore},
  {Dunham}, {Gautier}, {Geary}, {Gilliland}, {Gould}, {Howell}, {Jenkins},
  {Latham}, {Lissauer}, {Marcy}, {Rowe}, {Sasselov}, {Boss}, {Charbonneau},
  {Ciardi}, {Doyle}, {Dupree}, {Ford}, {Fortney}, {Holman}, {Seager},
  {Steffen}, {Tarter}, {Welsh}, {Allen}, {Buchhave}, {Christiansen}, {Clarke},
  {Das}, {D{\'e}sert}, {Endl}, {Fabrycky}, {Fressin}, {Haas}, {Horch},
  {Howard}, {Isaacson}, {Kjeldsen}, {Kolodziejczak}, {Kulesa}, {Li}, {Lucas},
  {Machalek}, {McCarthy}, {MacQueen}, {Meibom}, {Miquel}, {Prsa}, {Quinn},
  {Quintana}, {Ragozzine}, {Sherry}, {Shporer}, {Tenenbaum}, {Torres},
  {Twicken}, {Van Cleve}, {Walkowicz}, {Witteborn}, \& {Still}}]{2011BKB}
{Borucki}, W.~J., {Koch}, D.~G., {Basri}, G., {Batalha}, N., et al. 2011,
  \apj, 736, 19

\bibitem[{{Bowler} {et~al.}(2010){Bowler}, {Johnson}, {Marcy}, {Henry}, {Peek},
  {Fischer}, {Clubb}, {Liu}, {Reffert}, {Schwab}, \& {Lowe}}]{2010BJ}
{Bowler}, B.~P., {Johnson}, J.~A., {Marcy}, G.~W., {Henry}, G.~W., et al. 2010, \apj, 709, 396

\bibitem[{{Bromley} \& {Kenyon}(2011)}]{2011BK}
{Bromley}, B.~C., \& {Kenyon}, S.~J. 2011, \apj, 735, 29

\bibitem[{{Burke}(2008)}]{2008B}
{Burke}, C.~J. 2008, \apj, 679, 1566

\bibitem[{{Burke} {et~al.}(2007){Burke}, {McCullough}, {Valenti},
  {Johns-Krull}, {Janes}, {Heasley}, {Summers}, {Stys}, {Bissinger}, {Fleenor},
  {Foote}, {Garc{\'{\i}}a-Melendo}, {Gary}, {Howell}, {Mallia}, {Masi},
  {Taylor}, \& {Vanmunster}}]{2007BM}
{Burke}, C.~J., {McCullough}, P.~R., {Valenti}, J.~A., {Johns-Krull}, et al. 2007, \apj, 671, 2115

\bibitem[{{Carter} \& {Winn}(2009)}]{2009C}
{Carter}, J.~A., \& {Winn}, J.~N. 2009, \apj, 704, 51

\bibitem[{{Chib} \& {Greenberg}(1995)}]{1995C}
{Chib}, S., \& {Greenberg}, E. 1995, American Statistician, 49, 327

\bibitem[{{Dong} {et~al.}(2012){Dong}, {Katz}, \& {Socrates}}]{2012D}
{Dong}, S., {Katz}, B., \& {Socrates}, A. 2012, arXiv:1201.4399

\bibitem[Dressing et al.(2010)]{2010D} Dressing, C.~D., 
Spiegel, D.~S., Scharf, C.~A., Menou, K., 
\& Raymond, S.~N.\ 2010, \apj, 721, 1295 

\bibitem[{{Fabrycky} {et~al.}(2012){Fabrycky}, {Lissauer}, {Ragozzine}, {Rowe},
  {Agol}, {Barclay}, {Batalha}, {Borucki}, {Ciardi}, {Ford}, {Geary}, {Holman},
  {Jenkins}, {Li}, {Morehead}, {Shporer}, {Smith}, {Steffen}, \&
  {Still}}]{2012FL}
{Fabrycky}, D.~C., {Lissauer}, J.~J., {Ragozzine}, D., {Rowe}, J.~F., et al. 2012, arXiv:1202.6328

\bibitem[{{Fischer} {et~al.}(2005){Fischer}, {Laughlin}, {Butler}, {Marcy},
  {Johnson}, {Henry}, {Valenti}, {Vogt}, {Ammons}, {Robinson}, {Spear},
  {Strader}, {Driscoll}, {Fuller}, {Johnson}, {Manrao}, {McCarthy},
  {Mu{\~n}oz}, {Tah}, {Wright}, {Ida}, {Sato}, {Toyota}, \& {Minniti}}]{2005F}
{Fischer}, D.~A., {Laughlin}, G., {Butler}, P., {Marcy}, G., et al. 2005, \apj, 620, 481

\bibitem[{{Fischer} {et~al.}(2007){Fischer}, {Vogt}, {Marcy}, {Butler}, {Sato},
  {Henry}, {Robinson}, {Laughlin}, {Ida}, {Toyota}, {Omiya}, {Driscoll},
  {Takeda}, {Wright}, \& {Johnson}}]{2007F}
{Fischer}, D.~A., {Vogt}, S.~S., {Marcy}, G.~W., {Butler}, R.~P., et al. 2007, \apj, 669, 1336

\bibitem[{{Ford} \& {Rasio}(2006)}]{2006F}
{Ford}, E.~B., \& {Rasio}, F.~A. 2006, \apjl, 638, L45

\bibitem[{{Ford} \& {Rasio}(2008)}]{2008F}
---. 2008, \apj, 686, 621

\bibitem[{{Foreman-Mackey} {et~al.}(2012){Foreman-Mackey}, {Hogg}, {Lang}, \&
  {Goodman}}]{2012FD}
{Foreman-Mackey}, D., {Hogg}, D.~W., {Lang}, D., \& {Goodman}, J. 2012, arXiv:1202.3665

\bibitem[{{Gazak} {et~al.}(2011){Gazak}, {Johnson}, {Tonry}, {Eastman}, {Mann},
  \& {Agol}}]{2011G}
{Gazak}, J.~Z., {Johnson}, J.~A., {Tonry}, J., {Eastman}, J., et al. 2011, arXiv:1102.1036 

\bibitem[{{Gilliland} {et~al.}(2011){Gilliland}, {McCullough}, {Nelan},
  {Brown}, {Charbonneau}, {Nutzman}, {Christensen-Dalsgaard}, \&
  {Kjeldsen}}]{2011GM}
{Gilliland}, R.~L., {McCullough}, P.~R., {Nelan}, E.~P., {Brown}, et al. 2011, \apj, 726, 2

\bibitem[{{Goldreich} \& {Tremaine}(1980)}]{1980G}
{Goldreich}, P., \& {Tremaine}, S. 1980, \apj, 241, 425

\bibitem[Hansen(2010)]{2010H} Hansen, B.~M.~S.\ 2010, \apj, 
723, 285 

\bibitem[{{Hansen} \& {Murray}(2012)}]{2012H}
{Hansen}, B.~M.~S., \& {Murray}, N. 2012, \apj, 751, 158

\bibitem[{{Ida} \& {Lin}(2008)}]{2008I}
{Ida}, S., \& {Lin}, D.~N.~C. 2008, \apj, 673, 487

\bibitem[Jackson et al.(2008){\natexlab{a}}]{2008Ja} Jackson, B., Greenberg, 
R., \& Barnes, R.\ 2008{\natexlab{a}}, \apj, 678, 1396 

\bibitem[Jackson et al.(2008){\natexlab{b}}]{2008Jb}
---.\ 2008{\natexlab{b}}, \apj, 681, 1631 

\bibitem[{{Johnson} {et~al.}(2011){Johnson}, {Apps}, {Gazak}, {Crepp},
  {Crossfield}, {Howard}, {Marcy}, {Morton}, {Chubak}, \& {Isaacson}}]{2011JA}
{Johnson}, J.~A., {Apps}, K., {Gazak}, J.~Z., {Crepp}, et al. 2011, \apj, 730, 79

\bibitem[{{Jones} {et~al.}(2006){Jones}, {Butler}, {Tinney}, {Marcy}, {Carter},
  {Penny}, {McCarthy}, \& {Bailey}}]{2006J}
{Jones}, H.~R.~A., {Butler}, R.~P., {Tinney}, C.~G., {Marcy}, G.~W., et al. 2006, \mnras, 369, 249

\bibitem[{{Juri{\'c}} \& {Tremaine}(2008)}]{2008J}
{Juri{\'c}}, M., \& {Tremaine}, S. 2008, \apj, 686, 603

\bibitem[{{Kane} {et~al.}(2012){Kane}, {Ciardi}, {Gelino}, \& {von
  Braun}}]{2012KC}
{Kane}, S.~R., {Ciardi}, D.~R., {Gelino}, D.~M., \& {von Braun}, K. 2012, arXiv:1203.1631

\bibitem[{{Kane} \& {von Braun}(2009)}]{2009KV}
{Kane}, S.~R., \& {von Braun}, K. 2009, \pasp, 121, 1096

\bibitem[Kataria et al.(2011)]{2011K} Kataria, T., Showman, 
A.~P., Lewis, N.~K., et al.\ 2011, EPSC-DPS Joint Meeting 2011, 573 

\bibitem[{{Kipping}(2008)}]{2008K}
{Kipping}, D.~M. 2008, \mnras, 389, 1383

\bibitem[{{Kipping}(2010{\natexlab{a}})}]{2010K}
---. 2010{\natexlab{a}}, \mnras, 407, 301

\bibitem[{{Kipping}(2010{\natexlab{b}})}]{2010Kb}
---. 2010{\natexlab{b}}, \mnras, 408, 1758

\bibitem[{{Kipping} {et~al.}(2012){Kipping}, {Dunn}, {Jasinski}, \&
  {Manthri}}]{2012K}
{Kipping}, D.~M., {Dunn}, W.~R., {Jasinski}, J.~M., \& {Manthri}, V.~P. 2012,
  \mnras, 2413
  
  \bibitem[Kipping et al.(2009)]{2009K} Kipping, D.~M., Fossey, 
S.~J., \& Campanella, G.\ 2009, \mnras, 400, 398 

\bibitem[Kipping 
\& Tinetti(2010)]{2010KT} Kipping, D.~M., \& Tinetti, G.\ 2010, \mnras, 407, 2589 

\bibitem[Latham et al.(2011)]{2011LR} Latham, D.~W., Rowe, 
J.~F., Quinn, S.~N., et al.\ 2011, \apjl, 732, L24 

\bibitem[{{Lissauer} {et~al.}(2011){Lissauer}, {Ragozzine}, {Fabrycky},
  {Steffen}, {Ford}, {Jenkins}, {Shporer}, {Holman}, {Rowe}, {Quintana},
  {Batalha}, {Borucki}, {Bryson}, {Caldwell}, {Carter}, {Ciardi}, {Dunham},
  {Fortney}, {Gautier}, {Howell}, {Koch}, {Latham}, {Marcy}, {Morehead}, \&
  {Sasselov}}]{2011L}
{Lissauer}, J.~J., {Ragozzine}, D., {Fabrycky}, D.~C., {Steffen}, J.~H.,
  {Ford}, E.~B., et al. 2011,
  \apjs, 197, 8

\bibitem[{{Mandel} \& {Agol}(2002)}]{2002M}
{Mandel}, K., \& {Agol}, E. 2002, \apjl, 580, L171

\bibitem[Mann et al.(2012)]{2012MG} Mann, A.~W., Gaidos, E., 
L{\'e}pine, S., \& Hilton, E.~J.\ 2012, \apj, 753, 90 

\bibitem[Mardling(2007)]{2007M} Mardling, R.~A.\ 2007, 
\mnras, 382, 1768 

\bibitem[{{Moorhead} {et~al.}(2011){Moorhead}, {Ford}, {Morehead}, {Rowe},
  {Borucki}, {Batalha}, {Bryson}, {Caldwell}, {Fabrycky}, {Gautier}, {Koch},
  {Holman}, {Jenkins}, {Li}, {Lissauer}, {Lucas}, {Marcy}, {Quinn}, {Quintana},
  {Ragozzine}, {Shporer}, {Still}, \& {Torres}}]{2011M}
{Moorhead}, A.~V., {Ford}, E.~B., {Morehead}, R.~C., {Rowe}, J., et al. 2011, \apjs, 197, 1

\bibitem[Morton(2012)]{2012M} Morton, T.~D.\ 2012, 
arXiv:1206.1568 

\bibitem[Morton 
\& Johnson(2011)]{2011MJ} Morton, T.~D., \& Johnson, J.~A.\ 2011, \apj, 738, 170 

\bibitem[{{Naef} {et~al.}(2001){Naef}, {Latham}, {Mayor}, {Mazeh}, {Beuzit},
  {Drukier}, {Perrier-Bellet}, {Queloz}, {Sivan}, {Torres}, {Udry}, \&
  {Zucker}}]{2001N}
{Naef}, D., {Latham}, D.~W., {Mayor}, M., {Mazeh}, T., et al. 2001, \aap, 375, L27

\bibitem[{{Nagasawa} \& {Ida}(2011)}]{2011N}
{Nagasawa}, M., \& {Ida}, S. 2011, \apj, 742, 72

\bibitem[Naoz et al.(2012)]{2012NF} Naoz, S., Farr, W.~M., 
\& Rasio, F.~A.\ 2012, arXiv:1206.3529 

\bibitem[Nesvorny et al.(2012)]{2012N} Nesvorny, D., Kipping, 
D.~M., Buchhave, L.~A., et al.\ 2012, Science, 336, 1133 

\bibitem[{{O'Toole} {et~al.}(2009){O'Toole}, {Tinney}, {Jones}, {Butler},
  {Marcy}, {Carter}, \& {Bailey}}]{2009O}
{O'Toole}, S.~J., {Tinney}, C.~G., {Jones}, H.~R.~A., {Butler}, R.~P., et al. 2009, \mnras, 392, 641

\bibitem[{{Plavchan} {et~al.}(2012){Plavchan}, {Bilinski}, \& {Currie}}]{2012P}
{Plavchan}, P., {Bilinski}, C., \& {Currie}, T. 2012, arXiv:1203.1887

\bibitem[{{Ragozzine} \& {Holman}(2010)}]{2010R}
{Ragozzine}, D., \& {Holman}, M.~J. 2010, arXiv:1006.3727 

\bibitem[Santerne et al.(2012)]{2012SD} Santerne, A., 
D{\'{\i}}az, R.~F., Moutou, C., et al.\ 2012, arXiv:1206.0601 

\bibitem[{{Sing}(2010)}]{2010S}
{Sing}, D.~K. 2010, \aap, 510, A21

\bibitem[{{Socrates} {et~al.}(2012){Socrates}, {Katz}, {Dong}, \&
  {Tremaine}}]{2012SK}
{Socrates}, A., {Katz}, B., {Dong}, S., \& {Tremaine}, S. 2012, \apj, 750, 106

\bibitem[Spiegel et al.(2010)]{2010SR} Spiegel, D.~S., 
Raymond, S.~N., Dressing, C.~D., Scharf, C.~A., 
\& Mitchell, J.~L.\ 2010, \apj, 721, 1308 

\bibitem[{{Torres} {et~al.}(2010){Torres}, {Andersen}, \&
  {Gim{\'e}nez}}]{2010T}
{Torres}, G., {Andersen}, J., \& {Gim{\'e}nez}, A. 2010, \aapr, 18, 67

\bibitem[{{Valenti} \& {Fischer}(2005)}]{2005V}
{Valenti}, J.~A., \& {Fischer}, D.~A. 2005, \apjs, 159, 141

\bibitem[{{Valenti} \& {Piskunov}(1996)}]{1996V}
{Valenti}, J.~A., \& {Piskunov}, N. 1996, \aaps, 118, 595

\bibitem[{{Vogt} {et~al.}(1994){Vogt}, {Allen}, {Bigelow}, {Bresee}, {Brown},
  {Cantrall}, {Conrad}, {Couture}, {Delaney}, {Epps}, {Hilyard}, {Hilyard},
  {Horn}, {Jern}, {Kanto}, {Keane}, {Kibrick}, {Lewis}, {Osborne},
  {Pardeilhan}, {Pfister}, {Ricketts}, {Robinson}, {Stover}, {Tucker}, {Ward},
  \& {Wei}}]{1994V}
{Vogt}, S.~S., {Allen}, S.~L., {Bigelow}, B.~C., {Bresee}, L., et al. 1994, in Society
  of Photo-Optical Instrumentation Engineers (SPIE) Conference Series, Vol.
  2198, Society of Photo-Optical Instrumentation Engineers (SPIE) Conference
  Series, ed. {D.~L.~Crawford \& E.~R.~Craine}, 362

\bibitem[{{Ward}(1997)}]{1997W}
{Ward}, W.~R. 1997, \icarus, 126, 261

\bibitem[{{Winn}(2010)}]{2010W}
{Winn}, J.~N. 2010, {Exoplanet Transits and Occultations}, ed. {Seager, S.},
  55--77

\bibitem[{{Winn} {et~al.}(2009){Winn}, {Holman}, {Henry}, {Torres}, {Fischer},
  {Johnson}, {Marcy}, {Shporer}, \& {Mazeh}}]{2009W}
{Winn}, J.~N., {Holman}, M.~J., {Henry}, G.~W., {Torres}, G., et al. 2009, \apj,
  693, 794

\bibitem[{{Wright} {et~al.}(2011){Wright}, {Fakhouri}, {Marcy},
  {Han}, {Feng}, {Johnson}, {Howard}, {Fischer}, {Valenti}, {Anderson}, \&
  {Piskunov}}]{2011WF}
---. 2011{\natexlab{b}}, \pasp, 123, 412

\bibitem[{{Wu} \& {Lithwick}(2011)}]{2011W}
{Wu}, Y., \& {Lithwick}, Y. 2011, \apj, 735, 109

\bibitem[Wu 
\& Murray(2003)]{2003W} Wu, Y., \& Murray, N.\ 2003, \apj, 589, 605 


\end{thebibliography}

\end{document}